\documentclass[aps,prb,twocolumn,superscriptaddress,floatfix]{revtex4-2}
\usepackage[utf8]{inputenc}
\usepackage{natbib}
\usepackage[colorlinks=true,linkcolor=blue,citecolor=blue,urlcolor=blue]{hyperref}
\usepackage{graphicx}
\usepackage{latexsym}
\usepackage{amssymb}
\usepackage{amsmath}
\usepackage{amsfonts}
\usepackage{amsthm}
\usepackage{bm}
\usepackage{floatrow}
\usepackage[caption=false]{subfig}
\usepackage{bbm}
\usepackage{enumitem}
\usepackage{hyperref}
\usepackage{lipsum}
\usepackage{braket}
\usepackage{tikz}
\usepackage{pgfplots}
\usepackage{comment}
\usepackage{diagbox}
\usepackage{ifthen}
\usepackage{ragged2e}
\usepackage{multirow}
\usepackage[export]{adjustbox}
\usetikzlibrary{arrows,decorations.pathreplacing,decorations.markings,arrows.meta,patterns,3d}
\usepackage[normalem]{ulem} 
\usepackage{cancel}
\usepackage{microtype}
\usepackage{booktabs}

\theoremstyle{definition}

\pgfplotsset{compat=1.8}

\begin{document}

\title{Duality via Sequential Quantum Circuit in the Topological Holography Formalism}
\date{\today}

\author{Robijn Vanhove}
\affiliation{Department of Physics and Institute for Quantum Information and Matter, \mbox{California Institute of Technology, Pasadena, CA, 91125, USA}}
\affiliation{Department of Physics and Astronomy, Ghent university, Krijgslaan 281, S9, B-9000, Gent, Belgium}

\author{Vibhu Ravindran}
\affiliation{Department of Physics and Institute for Quantum Information and Matter, \mbox{California Institute of Technology, Pasadena, CA, 91125, USA}}

\author{David T. Stephen}
\affiliation{Department of Physics and Center for Theory of Quantum Matter, University of Colorado Boulder, Boulder, Colorado 80309 USA\looseness=-1}
\affiliation{Department of Physics and Institute for Quantum Information and Matter, \mbox{California Institute of Technology, Pasadena, CA, 91125, USA}}

\author{Xiao-Gang Wen}
\affiliation{Department of Physics, Massachusetts Institute of Technology,
Cambridge, Massachusetts 02139, USA}

\author{Xie Chen}
\affiliation{Walter Burke Institute for Theoretical Physics, \mbox{California Institute of Technology, Pasadena, CA, 91125, USA}}
\affiliation{Department of Physics and Institute for Quantum Information and Matter, \mbox{California Institute of Technology, Pasadena, CA, 91125, USA}}

\begin{abstract} 
Two quantum theories which look different but are secretly describing the same low-energy physics are said to be dual to each other. When realized in the Topological Holography formalism, duality corresponds to changing the gapped boundary condition on the top boundary of a topological field theory, which determines the symmetry of the system,  
while not affecting the bottom boundary where all the dynamics take place. In this paper, we show that duality in the Topological Holography formalism can be realized with a Sequential Quantum Circuit applied to the top boundary. As a consequence, the Hamiltonians before and after the duality mapping have exactly the same spectrum in the corresponding symmetry sectors, and the entanglement in the corresponding low-energy eigenstates differs by at most an area law term. These results reformulate the findings from [\href{https://arxiv.org/abs/2311.01439}{Lootens et al., arXiv:2311.01439 (2023)}] for dualities in $1+1D$ and extend them, using the Topological Holography framework, to higher dimensions.
\end{abstract}

\maketitle


Duality connects two different looking quantum theories and shows that they are secretly alike. Examples include the electro-magnetic duality of QED~\cite{jackson2021classical}, the Kramers-Wannier duality \cite{Kramers1941}, the Jordan Wigner transformation \cite{Jordan1928}, the particle-vortex duality \cite{Peskin1978,Dasgupta1981}, the boson-fermion duality \cite{Wilczek1982,Polyakov1988,Jain1989}, and (in a broader context) the AdS-CFT duality \cite{Maldacena1999}. The existence of dualities in these systems leads to strong constraints on their phases and phase transitions.

\begin{figure}[b]
    \centering
    \includegraphics[scale=0.45]{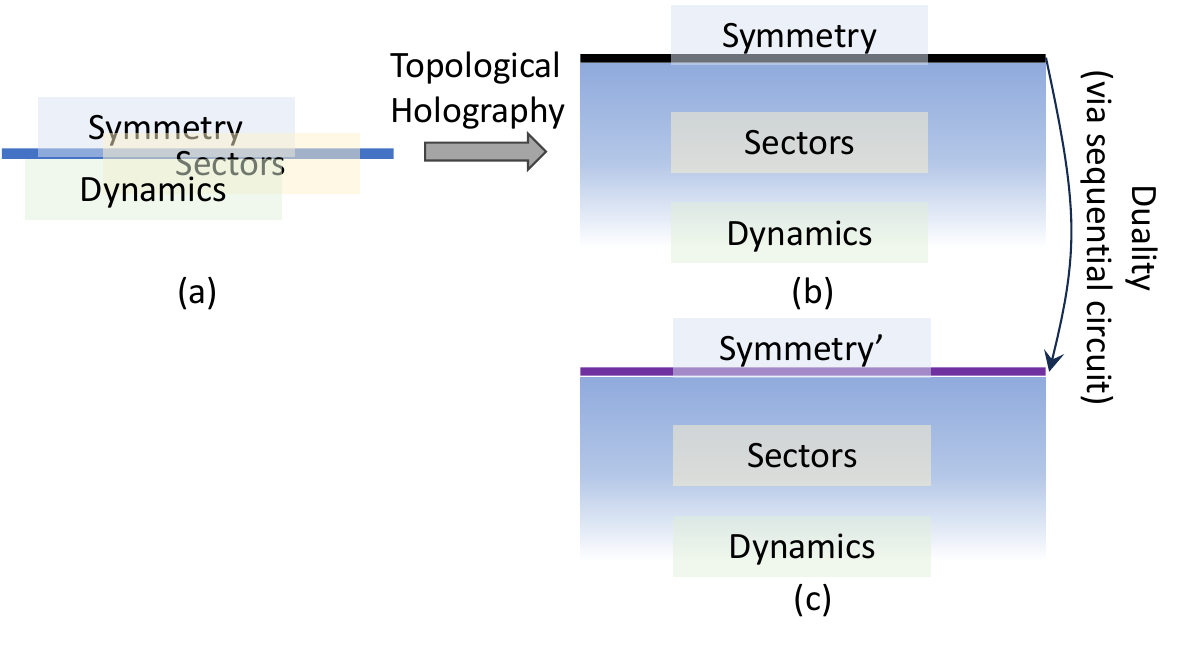}
    \caption{The Topological Holography formalism and duality. In this formalism, a $D+1$-dimensional system is realized as a `sandwich' structure with a $(D+1)+1$ dimensional topological bulk. The gapped top boundary determines the symmetry of the `sandwich', the bulk determines the (symmetry) sectors, and the bottom boundary contains all the (symmetric) dynamics. Duality is induced by the change of top boundary condition, which in our approach is implemented by a sequential quantum circuit acting near the top boundary.}
    \label{fig:TH}
\end{figure}

Duality takes on a very concrete meaning in the recently proposed Topological Holography (TH) / Symmetry Topological Field Theory (SymTFT) / Symmetry Topological Order (SymTO) formalism \cite{KZ150201690,Ji2020,Kong2020,Lichtman2021,Chatterjee2023symmetry,Moradi2022topological,Freed2023topological,lin2023asymptotic,kong2018gapless,kong2020mathematical,kong2021mathematical,kong2022one,kong2022categories,kong2024categories,xu2024categorical}. As shown in Fig.~\ref{fig:TH}, the Topological Holography formalism realizes a $D$ (spatial) dimensional quantum system as a `sandwich' structure \cite{KZ150201690,Freed2023topological}. The bulk of the sandwich is a $D+1$ dimensional topological system in its ground state with potential excitations described by a braided fusion higher category; the top edge of the sandwich is a gapped boundary of the topological bulk whose gapped excitations form a corresponding fusion higher category. The bulk together with the top boundary determines the low-energy Hilbert space of the `sandwich'. The top boundary specifies the symmetry of the quantum system represented by the sandwich while the bulk determines the symmetry sectors. The bottom edge, on the other hand, contains all the dynamical terms of the system.
The composite system simulates the low-energy physics of a $D$-dimensional quantum system when the bulk gap and the top-boundary gap approaches infinity.
For example, if the $2D$ bulk has $\mathbb{Z}_2$ topological order and the top boundary is the smooth boundary of flux condensation, the sandwich realizes a $1D$ chain with global $\mathbb{Z}_2$ symmetry with no anomaly. 
SymTO and SymTFT are similar constructions with minor differences. SymTO emphasizes more on the lattice model perspective while SymTFT more on the field theory perspective. In particular, in the sandwich structure described above, the lattice realization of the bulk in SymTO can contain perturbations that break any symmetry (for example the $e$-$m$ exchange symmetry of the toric code) while the field theory description of the bulk in SymTFT preserves those symmetries. Notably, the notion of Topological Holography has appeared in specific lattice constructions, see for example Ref.~\onlinecite{you2014wave,aasen2016topological,aasen2020topological,vanhove2018mapping,vanhove2022topological} in the context of $2D$ (critical) satistical mechanics models. Moreover, topological holography, and similar ideas as the ones depicted in Fig.~\ref{fig:TH} have appeared in the mathematical physics literature under the name `Topological Wick rotation' \cite{kong2018gapless, kong2020mathematical, kong2021mathematical}.

The holographic picture gives us a very general and unified
understanding of duality: a duality transformation is induced by a change in the
top boundary condition, while the
bulk and the bottom boundary of the sandwich remain invariant \cite{Kong2020,
Chatterjee2023holographic,Huang2023topological}. Since the gapped excitations on the top boundary (described by a fusion higher category) may change, the
systems related by the duality transformation may not have the same symmetry.
For example, if the bulk has the $3D$ $\mathbb{Z}_2$ topological order, changing from the charge condensed boundary to the flux condensed boundary corresponds to coupling $2D$ systems with 0-form $\mathbb{Z}_2$ symmetry with a dynamical $\mathbb{Z}_2$ gauge field and results in a gauge theory with 1-form $\mathbb{Z}_2$ symmetry. 

But how can two systems with different symmetries be similar? Through some simple examples, it was conjectured \cite{Ji2020,Kong2020} that if we
restrict ourselves to the symmetric sub-Hilbert spaces of the respective
symmetries, the two systems related by the duality transformation are identical. This is because the symmetric sub-Hilbert space is described by dynamical operators near the bottom boundary, which is not changed under the duality transformation. In fact, such a correspondence can be extended to sub-Hilbert space sectors with nontrivial charge and flux as well as long as they are matched properly in the two systems before and after the duality mapping. The holographic picture provides a simple way to match the sectors since they are labeled by logical operators in the topological bulk, and the bulk is not changed either under the duality transformation. Therefore, as the duality transformation maps one Hamiltonian to another Hamiltonian with a different symmetry, it implements a unitary transformation between corresponding symmetry sectors of the two systems.

In some cases, the fusion higher
category describing the gapped excitations on the two top boundaries is the same, so the systems before and after duality have the same symmetry. With additional fine-tuning, the Hamiltonian can be invariant under duality, in which case the duality becomes a self-duality and the transformation can be regarded as a kind of generalized symmetry. For example, the $2D$ $\mathbb{Z}_2$ topological order discussed above has an
electric-magnetic automorphism. Under this automorphism, the flux-condensed top
boundary maps to the charge-condensed top boundary. The whole sandwich still
realizes a $1D$ system with $\mathbb{Z}_2$ symmetry. This is the well known
Kramers-Wannier duality of $1D$ Ising chain, which maps an Ising Hamiltomian $H
= - \sum J Z_iZ_{i+1} + h X_i$ to $\tilde H = - \sum h Z_iZ_{i+1} + J X_i$.
When $J=h$, the model is self-dual -- invariant under the additional ``duality symmetry''. To find the holographic description of the enhanced symmetry, we make the electric-magnetic autmorphism
into an electric-magnetic exchange symmetry of the bulk topological order. After gauging
the electric-magnetic exchange symmetry, we obtain a new bulk topological order -- the double-Ising topological order, which describes the enhanced symmetry of the self-dual Ising model. Note that our approach to duality in this paper -- sequential quantum circuit applied to the top boundary of the sandwich -- treats self-duality and the more general type of duality (where symmetries can change) on equal footing and all results derived thereby apply to general dualities in the TH formalism.

What changes and what remains invariant under duality transformation? In this paper, we give an explicit answer to this question using the Sequential Quantum Circuit (SQC). We find that, for the two quantum systems related by duality, their phase diagrams have the same structure and duality maps between the corresponding phases / phase transitions. For two symmetric models (one from each system) related by duality, common wisdom is that the low-energy physics of the two models are the same. In this paper, we show explicitly that the low-energy spectrum of the models are exactly the same within the respective symmetric sub-Hilbert space with no symmetry defect. Also, the low-energy spectrum of the model remains exactly the same
when all the symmetry sectors of different symmetry charges/defects are included (or if sectors are matched in the proper way). Moreover, the entanglement in the low energy eigenstates can change by at most an area law term. We show these results as a consequence of the fact that duality in the TH formalism can be realized by applying a Sequential Quantum Circuit (SQC) to the top boundary and changing it from one gapped state to another. The SQC was introduced in Ref.~\onlinecite{Chen2024} as maps between different gapped phases. It was defined as quantum circuits where each local degree of freedom is acted upon by only a finite number of gates. As a consequence, an SQC changes the entanglement in a quantum state by at most an area law term. The SQC used in this paper to induce duality transformation in the TH formalism is a unitary transformation within each symmetry sector, therefore completely preserving the spectrum in each sector. Moreover, since the SQC changes the entanglement in a quantum state by at most an area law term, it preserves the entanglement area law and all terms more significant than it. A similar result was obtained in Ref.~\onlinecite{Lootens2023lowdepth} in $1+1D$, building on the unifying framework for dualities with general categorical symmetries developed in Refs.~\onlinecite{Lootens2023, Lootens2024dualities} in terms of Matrix Product Operator intertwiners and the module category formalism. Dualities in the Topological Holography framework have been discussed in terms of the partition function of the system in Ref.~\onlinecite{Huang2023topological}.
 
The advantage of our approach is two-fold. First, our circuits for implementing dualities have a simple and intuitive construction that comes from existing circuits for generating topological phases \cite{Chen2024}. Second, the fact that operators in the bulk and two boundaries play distinct roles in a TH sandwich (Fig.~\ref{fig:TH}), combined with the fact that our circuits act only near the top boundary of the sandwich, clearly delineates what features of a quantum system can and cannot change under duality.

We need to point out that our statements do not have direct implications for dualities between field theories. In particular, our formalism does not provide extra information about which field theory a particular quantum model flows to in the IR. Therefore, it does not directly address open questions about dualities between field theories. If certain field theories can be shown to describe the IR theory of certain points in the phase diagram (like the Ising critical point in the transverse field Ising model), then our conclusion applies to the spectrum and the low-energy eigenstates of the field theories.

The paper consists of three main sections and is structured as follows. In section \ref{sec:2DTC}, we review the Kramers-Wannier duality in the $1+1D$ transverse field Ising model and show explicitly how the duality is a unitary mapping between all the sectors combined. We show how the Kramers-Wannier duality and the sector correspondence is recovered in the TH picture through a sequential quantum circuit that changes the top boundary. Lastly, we discuss the Jordan-Wigner transformation in the TH setup by writing down a sequential quantum circuit that changes the top boundary of the toric code sandwich to the fermionic boundary (where the fermionic bulk quasiparticles can condense). In section \ref{sec:SN}, we generalize the above to $2+1D$ TH constructions where the bulk is a general string-net, which allows for dualities between models with general non-invertible (categorical) symmetries. We show how the sector re-interpretation is more subtle in the general case and we write down isometric sequential quantum circuits that change between different gapped boundaries. We illustrate the above for the dualities associated to the sandwich with the bulk the $S_3$ quantum double, which hosts gapped boundaries for which the sandwich symmetry is Rep($G$). Finally, in section \ref{sec:3DTC}, we showcase duality in a $2+1D$ sandwich with the $3+1D$ toric code in the bulk by constructing two-dimensional sequential quantum circuits that map between the gapped boundaries of the $3+1D$ toric code. The associated dualities relate $2+1D$ systems with a $0$-form symmetry to their duals with $1$-form symmetries.

\section{Toric Code and Kramers-Wannier Duality}
\label{sec:2DTC}
In this section we will discuss the simplest example of duality in the TH formalism: the Kramers-Wannier self-duality in $1+1D$ spin chains with global $\mathbb{Z}_2$ symmetry. First in section~\ref{sec:KW}, we review how the duality is usually discussed in terms of the mapping of operators in the spin-$1/2$ chain. We emphasize on the splitting of the Hilbert space into sectors according to symmetry charge and symmetry twisted boundary conditions and on how the duality permutes these sectors. Next in section~\ref{sec:KW_TH}, we realize this model as a TH sandwich, i.e. as a $1+1D$ strip with the $2+1D$ toric code in the bulk and an anyon condensed gapped top boundary. In the sandwich construction, the duality is realized by changing the top boundary from one type of condensate to another and permuting the symmetry charge and symmetry twist sectors which are now labeled by bulk anyons. In section~\ref{sec:KW_SQC}, we show how the change in top boundary condition can be achieved with a $1D$ sequential quantum circuit applied to the top boundary without affecting the bulk or the bottom boundary. This shows that the low-energy spectrum within each sector is preserved. Finally, we illustrate in section~\ref{sec:JW} how the Jordan Wigner transformation naturally arises in the same sandwich by changing to a `fermion condensed' boundary.

\subsection{Review: Kramers-Wannier Duality} 
\label{sec:KW}

In this section, we review the Kramers-Wannier self duality of the transverse field Ising model of the spin $1/2$ chain. We explicitly compare the low-energy (ground and first excited) spectrum before and after the mapping and see how they match. This is a feature we will come back to from the TH point of view.

Consider the transverse field Ising model (TFIM) on a $1D$ lattice with $N$ spin $1/2$ degrees of freedom and periodic boundary condition: 
\begin{equation}
H_{\text{TFI}} = -\sum_{i=1}^N\left( Z_iZ_{i+1} + gX_i\right)\label{eq:H_TFI}
\end{equation}
Here the indices are taken mod $N$. There is a global $\mathbb{Z}_2$ spin flip symmetry realized as $\prod_i X_i$ which can either be broken ($|g|<1$) or preserved ($|g|>1$). The Kramers-Wannier duality is given by the following operator mapping: 
\begin{equation}
    \begin{aligned}
    X_i &\to  Z_iZ_{i+1}\\
    Z_iZ_{i+1} &\to X_{i+1} \\
    \end{aligned}
    \label{eq:KW_map}
\end{equation}
This maps local symmetric operators to local symmetric operators while keeping the local operator algebra (commutation relation among local operators) invariant. But globally, there is a problem with this. The $\mathbb{Z}_2$ symmetry operator is mapped to identity while the new $\mathbb{Z}_2$ symmetry operator is mapped from identity
\begin{equation}
    \begin{aligned}
    \prod_{i=1}^N X_i &\to \prod_{i=1}^N Z_iZ_{i+1} = 1\\
    1 = \prod_{i=1}^N Z_iZ_{i+1} &\to \prod_{i=1}^N X_i
    \end{aligned}
\end{equation}
If we are restricted to the no-charge sector of the global symmetry ($\prod_i X_i = 1$) and periodic boundary condition, the duality mapping is well defined. In this case, we have a self-duality: the Hamiltonian terms in Eq.~\ref{eq:H_TFI} are exchanged ($Z_iZ_{i+1} + gX_i \to X_i + g Z_iZ_{i+1}$), mapping the model back to itself with a parameter change $g\to 1/g$. 

In order to extend this mapping to symmetry charged states, we need to allow the symmetry-twisted (anti-periodic) boundary condition. 
Modify the Hamiltonian to 
\begin{equation}
    H = -\sum_{i=1}^{N-1}\left(Z_iZ_{i+1} + gX_i\right) - \left(\eta Z_NZ_1 + gX_N\right)
\end{equation}
where $\eta = \pm1$ labels the symmetry twist (boundary condition). By adding this parameter, we have doubled the total Hilbert space we are considering. Each of these spaces splits into symmetry charged and uncharged states and the total Hilbert space is a direct sum of these four sectors. Now we can alter the operator mapping so that it is well defined for $\mathbb{Z}_2$ odd states. Following \cite{radicevic2019spin}, and calling the symmetry twists $\eta$ and  $\eta' = \pm 1$ on the two sides of the duality, we achieve this by modifying the last term in the mapping:
\begin{equation}
    \begin{aligned}
        X_N &\to \eta' Z_NZ_1\\
        \eta Z_NZ_1 &\to X_1\\
    \end{aligned}
\end{equation}

This transformation allows symmetry charged states on both sides of the duality: 
\begin{equation}
    \begin{aligned}
     \prod_{i=1}^N X_i &\to \eta'\\
    \eta &\to \prod_{i=1}^N X_i\\  
    \end{aligned}
\end{equation}
 The above equations say that under the duality, the four sectors are permuted amongst themselves, with the symmetry charge and symmetry twist labels getting swapped. When we consider the mapping from one sector to the corresponding sector or between all four-sectors combined, the duality is a unitary (isometric) mapping.

The unitarity of the duality is reflected in the matching of the spectrum of the corresponding sectors before and after the mapping. Table~\ref{KWspectrum} lists the ground and excited states in the corresponding sectors before and after the duality. We consider the fixed-point models for simplicity in both the symmetry-breaking phase ($g=0$) and the symmetric phase ($g\to +\infty$). As mentioned above, we label the Hilbert space sectors by (symmetry charge, boundary condition), so the $(1,1)$ and $(-1,-1)$ sectors are mapped to themselves and the $(-1,1)$ and $(1,-1)$ are exchanged by the duality. 

\begin{table}[]
\centering
\begin{tabular}{|c|c|c|} 
    \hline 
    & Symmetry breaking & Symmetric\\
    \hline
    Sector & $(1,1)$ & $(1,1)$ \\ \hline
    GS & $\ket{\psi_{sb}} = \ket{00...0} + \ket{11...1}$ & $\ket{\psi_{sym}} = \ket{++...+}$ \\ \hline
    Ex & $\prod_{i_0}^{i_1} X_i\ket{\psi_{sb}}$ & $Z_{i_0}Z_{i_1} \ket{\psi_{sym}}$ \\
     \hline \hline
     Sector & $(1,-1)$ & $(-1,1)$ \\ \hline
    GS & $\prod_{i_0}^{N} X_i\ket{\psi_{sb}}$ & $Z_{i_0}\ket{\psi_{sym}}$\\ \hline
    Ex & $\prod_{i_1}^{i_2}X_i \prod_{i_0}^{N} X_j\ket{\psi_{sb}}$ & $Z_{i_0}Z_{i_1}Z_{i_2} \ket{\psi_{sym}}$ \\ 
    \hline \hline
     Sector & $(-1,1)$ & $(1,-1)$ \\ \hline
    GS & $\ket{\psi_{sb}}' = \ket{00...0} - \ket{11...1}$ & $\ket{\psi_{sym}}$\\ \hline
    Ex & $\prod_{i_0}^{i_1} X_i\ket{\psi_{sb}}$ & $Z_{i_0}Z_{i_1}\ket{\psi_{sym}}$ \\
    \hline \hline
     Sector & $(-1,-1)$ & $(-1,-1)$ \\ \hline
    GS & $\prod_{i_0}^{N} X_i\ket{\psi_{sb}}'$ & $Z_{i_0}\ket{\psi_{sym}}$\\ \hline
    Ex & $\prod_{i_1}^{i_2}X_i \prod_{i_0}^{N} X_j\ket{\psi_{sb}}'$ & $Z_{i_0}Z_{i_1}Z_{i_2}\ket{\psi_{sym}}$ \\ \hline
\end{tabular}
\caption{The matching of the low-energy spectrum in the corresponding symmetry sectors before and after the Kramers-Wannier duality. The ground states (GS) and first excited states (Ex) at the fixed point of the symmetry-breaking phase and the symmetric phase are listed. It can be checked that the number of ground states and first excited states in the two phases are the same in the corresponding sectors.}
\label{KWspectrum}
\end{table}

The $(1,1)$ sector undergoes the standard Kramers-Wannier mapping of Eq.~\ref{eq:KW_map} sending the symmetry breaking ground state with $+1$ $\mathbb{Z}_2$ charge $|00\dots0\rangle + |11\dots1\rangle$ to the symmetric state $|++\dots+\rangle$. Excitations above the former state are domain walls, of which there are $\binom{N}{2}$ (we can choose the start and endpoint arbitrarily). On the other side (the symmetric phase), the excitations are states with two $-$ spins of which there are also $\binom{N}{2}$. Similarly, there are $N$ $(1,-1)$ ground states in the symmetry breaking phase and correspondingly $N$ $(-1,1)$ ground states in the symmetric phase. The excited states in the latter are states with three spins flipped $-$ of which there are $\binom{N}{3}$. The excited states in the former are states with two pairs of domain walls, one starting at the $N$th site. Counting these also gives $\binom{N}{3}$ as expected. The counting in the other two cases are similar and are shown in table~\ref{KWspectrum} for completeness. The matching of the low-energy spectrum between the two sides is a general feature guaranteed by the unitarity of the duality mapping between the corresponding sectors, which can be implemented as a sequential quantum circuit in the $1D$ chain \cite{Chen2024,Seifnashri2024arxiv}.

%
\subsection{KW Duality in the TH formalism}
\label{sec:KW_TH}

\begin{figure}
    \centering
    \includegraphics[scale=0.5,page=51]{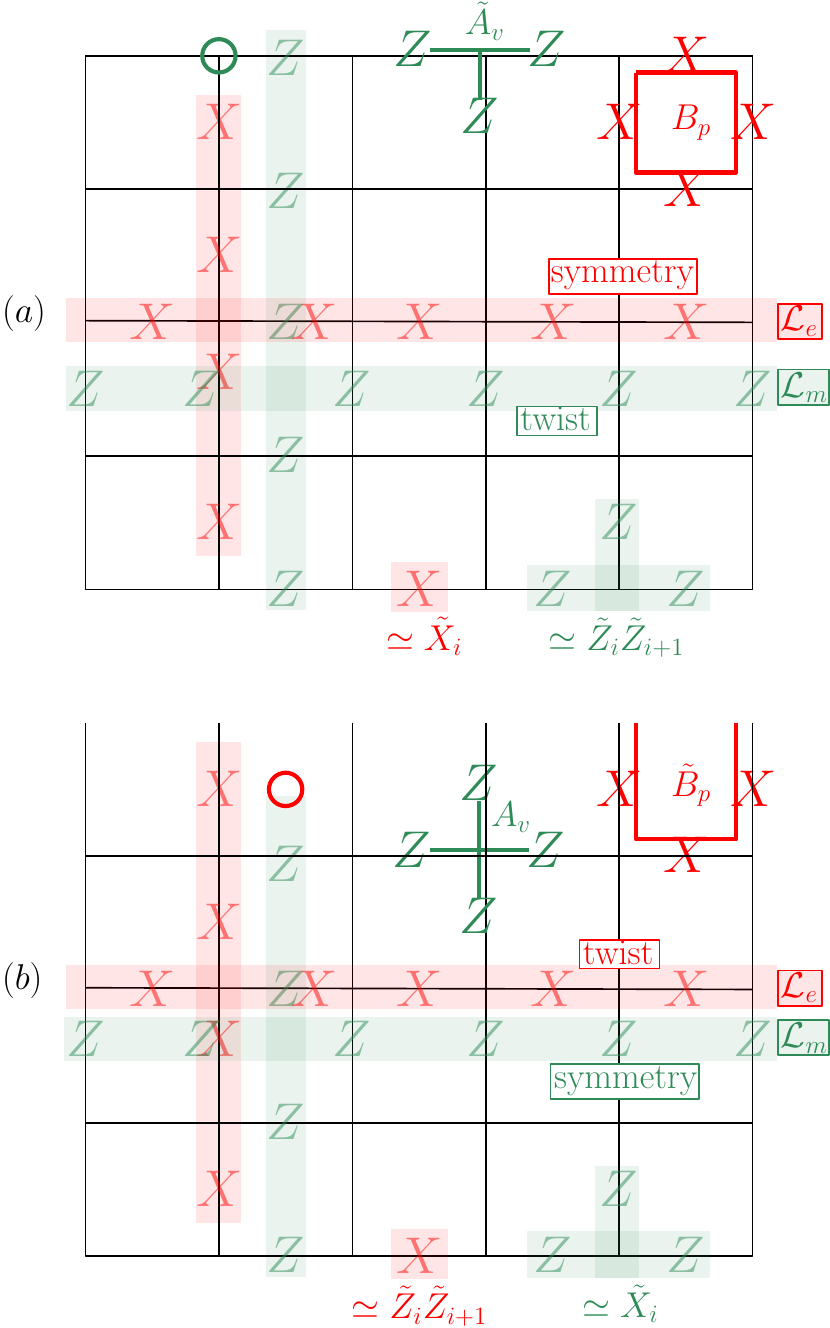}
    \caption{Toric code in the sandwich picture with (a) smooth top boundary and (b) rough top boundary. In addition to all bulk stabilizers (Eq.~\ref{TCHam}), we enforce additional stabilizers depending on the top boundary (unshaded terms). The hollow circles denote the locations of the omitted boundary stabilizers. The eigenvalues of the horizontal strings denote the symmetry and twist sectors, while the vertical strings toggle between these sectors. The single $X$ and $Z^{\otimes 3}$ on the bottom boundary are identified with the symmetric operator algebra of the effective $1D$ chain as indicated.}
    \label{fig:TCTH}
\end{figure}
In order to make contact with the TH formalism, we realize the TFIM as a $1+1D$ sandwich with the $2+1D$ Toric Code in the bulk. In this section we describe this identification explicitly and see how the Kramers-Wannier duality arises naturally in this framework as a change of top boundary condition of the sandwich. 

Our TH setup is shown in Fig.~\ref{fig:TCTH}. Taking the following Hamiltonian for the bulk toric code.
\begin{align} \label{TCHam}
 H_{\text{TC}}^{\text{bulk}} = -\sum_{v} \left[\includegraphics[valign=c,page=21,scale=0.35]{duality_figures}\right] - \sum_{p} \left[\includegraphics[valign=c,page=22,scale=0.35]{duality_figures}\right],
\end{align}
A flux excitation ($\bm{m}$) violates the red plaquette term while a charge excitation ($\bm{e}$) violates the green vertex term. Recall that we can identify two possible gapped boundary conditions~\cite{Bravyi1998arxiv}: a flux ($\bm{m}$)-condensed boundary, called `smooth', and a charge ($\bm{e}$)-condensed boundary, called `rough' (we write bulk anyons in bold). Concretely, the two boundaries correspond to ending the lattice on horizontal and vertical edges respectively. 
We consider a toric code ground state living on a cylinder (periodic boundary conditions in the horizontal direction) and fix the top boundary to be smooth or rough. 
In our TH picture, the bottom boundary is not fixed but left free, leaving dangling degrees of freedom in order to represent the dynamics of our effective $1+1D$ model.

There are several kinds of operators in the formalism that each play a different role, as pictured in Fig.~\ref{fig:TCTH}. First, there are the local stabilizers whose $+1$ eigenspace specifies the low-energy subspace in which the corresponding $1+1D$ theory is defined. These consist of all bulk stabilizers, as defined in Eq.~\ref{TCHam}, and truncated vertex (for a smooth boundary) or plaquette (for a rough boundary) stabilizers at the top boundary. Importantly, we omit a single boundary term from the set of stabilizers in order to accommodate twist sectors. The low energy subspace specified by the stabilizers then splits into four sectors. These sectors are labelled by the $\pm 1$ eigenvalues of the horizontal $X$ and $Z$ strings winding around the cylinder on the direct and dual lattice, respectively. In the remainder of this paper, we will label the horizontal operators (Wilson lines in the periodic direction) as $\mathcal{L}_{\alpha}$ (using Greek letter for bulk anyons). For the toric code, this corresponds to:
\begin{align*}
\mathcal{L}_{\bm{e}} &= \prod_{e \in C} X_{e} \\
\mathcal{L}_{\bm{m}} &= \prod_{e'\in C'} Z_{e'},
\end{align*}
with $e\in C$ ($e\in C'$) the edges along a non-contratible path on the direct (dual) lattice in the horizontal direction (Fig. \ref{fig:TCTH}).
In this paper, we only consider states of the sandwich which have definite eigenvalues of these strings, as these will correspond to states living in definite symmetry charge and twist sectors for Abelian theories such as the toric code. The operators that toggle between different sectors are vertical $X$ and $Z$ strings. On the smooth (rough) boundary, the vertical $X$ ($Z$) string excites a boundary stabilizer unless it ends at the vertex (plaquette) corresponding to the omitted term, whereas the vertical $Z$ ($X$) string can end anywhere on the top without creating any excitations. Finally, there are two kinds of local operators at the bottom boundary---a single $X$ and three $Z$'s as depicted in Fig.~\ref{fig:TCTH}---that commute with all the aforementioned operators, and therefore act only within a given sector of the low energy subspace. These form a complete set of symmetric operators for each sector and determine the dynamics of the $1+1D$ theory.

Now we identify the operators in the sandwich picture with those in the $1+1D$ TFIM, which we label as $\tilde{X}_i$ and $\tilde{Z}_i\tilde{Z}_{i+1}$. With a smooth boundary, we call the horizontal $X$ string the $\mathbb{Z}_2$ symmetry operator since its corresponding charged operator, the vertical $Z$ string, does not create any excitations on the top boundary. The eigenvalue of the horizontal $Z$ string then determines the twist sector. With these identifications, the $Z^{\otimes 3}$ term on the bottom boundary is equivalent to the product of two neighboring vertical $Z$ strings up to bulk stabilizers, so it corresponds to the Ising term $\tilde{Z}_i\tilde{Z}_{i+1}$. The single $X$ on the bottom boundary corresponds to the transverse field $\tilde{X}_i$ of the TFIM due to its commutation relation with the $Z^{\otimes 3}$ terms and a product of all such $X$'s is equivalent to the symmetry operator in the low-energy subspace. When we enforce all $X$ terms on the bottom boundary to have an eigenvalue $+1$, this makes a rough bottom boundary. Alongside a smooth top boundary, this corresponds to the symmetric phase of the TFIM. Similarly, setting all $Z^{\otimes 3}$ equal to $+1$ creates a smooth bottom boundary and the whole sandwich corresponds to the symmetry-breaking phase. If, instead, the top boundary is set to be rough, all roles are reversed. The horizontal $Z$ string is now the $\mathbb{Z}_2$ symmetry, the $X$ term on the bottom boundary is identified with $\tilde{Z}_i\tilde{Z}_{i+1}$, while the $Z^{\otimes 3}$ term is identified with $\tilde{X}_i$, and a rough (smooth) bottom boundary corresponds to the symmetry breaking (symmetric) phase of the Ising chain.

Finally, we describe the duality that arises when changing the top boundary from smooth to rough. From the above, we see that the change is equivalent to exchanging the transverse field and the Ising terms of the TFIM, thereby exchanging the symmetric and symmetry-breaking phases, which reproduces the KW duality. Furthermore, since the symmetry charge and symmetry twist sectors are labeled by the horizontal bulk string operators that are not affected by the change of boundary, these operators and their eigenvalues remain unchanged (this will be made more explicit in the following section). However, the interpretation of the string operators as labeling either symmetry charge or twist sector switches, such that the corresponding sector labels are exchanged. This matches the sector mapping of Kramers-Wannier as reviewed in the previous section.

\subsection{KW Duality via Sequential Quantum Circuit}
\label{sec:KW_SQC}

We now show how to implement the KW duality in the TH picture by constructing a sequential quantum circuit that changes the top boundary from rough to smooth. The circuit acts only along the top boundary without any action in the bulk or at the bottom boundary.

\begin{figure}[t]
\includegraphics[scale=0.55, page=52]{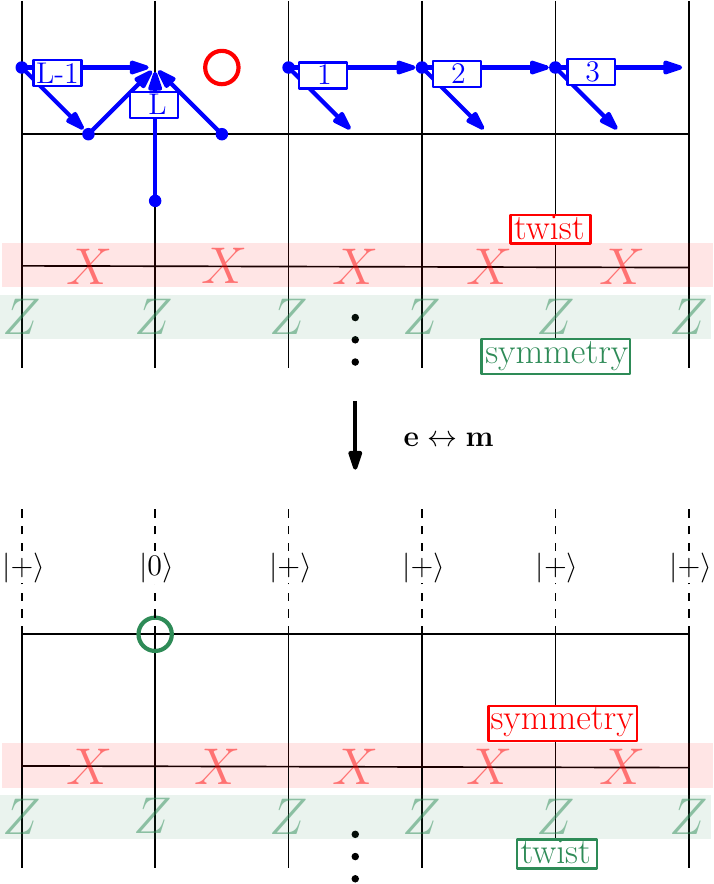}
\caption{Kramers-Wannier duality via unitary sequential quantum circuit in the toric code sandwich going from rough to smooth boundary. After the circuit, we are left with a smooth boundary with an omitted boundary stabilizer at the position marked by the hollow circle, and all rough legs are disentangled into product states. After the circuit, the eigenvalues of the bulk symmetry and twist operators remain unchanged, but their roles are reversed.}
\label{KWseqCirc}
\end{figure}

Quantum circuits that change boundary types in $2D$ topological orders can be constructed by adapting the sequential quantum circuits defined in Ref.~\onlinecite{Chen2024} (see section \ref{sec:SNSQC}). The circuits act near the boundary and are of linear depth in the periodic direction. For the toric code, the circuit is shown in Fig.~\ref{KWseqCirc}. It takes a simple form, consisting of a series of controlled-$X$ (CNOT) gates (blue arrows).

Let's look at what the circuit does to the stabilizers at the top boundary when it change from rough to smooth. The boundary plaquette terms and bulk vertex terms near the boundary (except those near the point where the boundary stabilizer is omitted) are mapped as follows:
\begin{align} \label{rough-smooth}
\begin{split}
\includegraphics[valign=c,page=25,scale=0.5]{duality_figures}\\
\includegraphics[valign=c,page=26,scale=0.5]{duality_figures}
\end{split}
\end{align}
where the CNOT gates are centered on the top-right plaquette in the lower expression. The vertex term near the omitted stabilizer is mapped as,
\begin{equation}
    \includegraphics[valign=c,page=55,scale=0.5]{duality_figures}
\end{equation}

We see that the rough edges get disentangled from the bulk by the circuit, and that the new boundary stabilizers are that of the smooth boundary (with one boundary term omitted). Since the horizontal string operators in the bulk are not touched by the circuit, their eigenvalues remain unchanged, as claimed in the previous section. To map from smooth to rough, we can do the same process in reverse. We initialize new qubits on vertical edges above the smooth boundary according to Fig.~\ref{KWseqCirc} and apply the same gates in reverse order.

The fact that the change of the top boundary condition in the sandwich can be achieved with a unitary quantum circuit acting only along the top boundary corroborates some of the features of the Kramers-Wannier duality discussed above. In particular, due to the unitarity of the mapping, the spectrum in corresponding sectors before and after the duality are exactly the same -- the eigenstates have the same energy and the same degeneracy. Moreover, since the sequential quantum circuit preserves the entanglement area law of the wave-function it acts on, the entanglement in the wave-function before and after the duality transformation can only differ by an area law term (a constant in $1D$). This conclusion applies to the self-dual Ising critical point as well when eigenstates in different sectors are related by the duality.

\subsection{Jordan-Wigner transformation}
\label{sec:JW}

\begin{figure}
    \centering
    \includegraphics[scale=0.55,page=54]{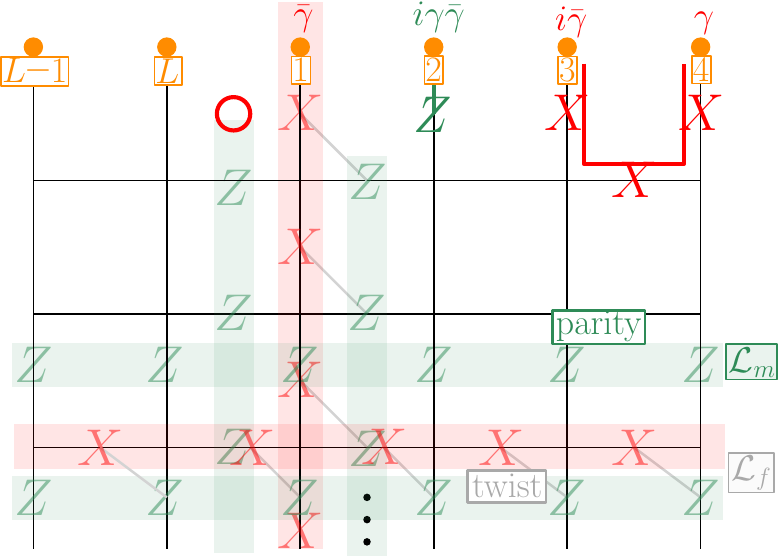}
    \caption{Toric code with fermionic, $\bm{em}$($\bm{f}$)-condensed, boundary. The orange disks represent fermions with Majorana operators $\gamma$ and $\bar{\gamma}$. $Z$s on the rough edges are dressed with a parity operator on the fermionic degrees of freedom and the boundary plaquette stabilizers are dressed with hopping operators (the bulk Hamiltonian is still as in Eq.~\ref{TCHam}). The vertical $\bm{f}$-string is dressed with $\bar{\gamma}$ and commutes with all the stabilizers. The total fermion parity is represented in the bulk by $\mathcal{L}_{\bm{m}}$. We omit one boundary plaquette stabilizer, allowing for twisted boundary conditions in the fermionic sandwich. The boundary condition is determined by the eigenvalue of $\mathcal{L}_{\bm{f}}$.}
    \label{fig:TCFermion}
\end{figure}

It was recently shown how the TH framework can be extended to fermionic boundaries \cite{wen2024topological, bhardwaj2024fermionic, huang2024fermionic}. In this section, we show how the Jordan-Wigner transformation can be obtained in the TH formalism by changing the top boundary of the toric code to the fermionic boundary by a sequential quantum circuit. The boundary change takes the transverse field Ising model to free fermions. The fermionic boundary in the toric code is obtained by introducing physical fermions on the boundary, allowing for $\bm{em}$ ($\bm{f}$) fermionic quasiparticles from the bulk to condense by coupling them to the physical fermions on the boundary. It is useful to work with Majorana operators for the fermion (on site $n$): $\gamma_n = c_n + c_n^{\dagger}$ and $\bar{\gamma}_n = i(c_n - c_n^{\dagger})$, where $c^{\dagger}$ and $c$ are the fermion creation and annihilation operators. The toric code TH with fermionic boundary is shown in Fig.~\ref{fig:TCFermion}. The boundary can be viewed as a rough boundary with an extra fermion on each rough leg. The fermions are coupled to the spins by edge terms and boundary plaquette terms. The edge term consists of a parity operator on the fermion $i\gamma\bar{\gamma}$ ($=1-2c^{\dagger}c$) and a $Z$ on the rough leg and the plaquette term of the rough boundary is coupled to a nearest-neighbor hopping-pairing term $i\bar{\gamma}_n\gamma_{n+1}$. We have labeled the fermionic sites in Fig. \ref{fig:TCFermion} in order to consistently evaluate the action of the operators in the remainder of this section. Using the anti-commutation relations of the Majorana operator on different sites and $(\gamma_n)^2 =(\bar{\gamma}_n)^2=1$, one can check that a vertical $\bm{f}$-string, dressed with $\bar{\gamma}_n$ on the boundary fermion, commutes with all the boundary stabilizers. We omit one boundary plaquette stabilizer to allow for twisted boundary conditions in the fermionic sandwich. Using the boundary stabilizers, one can check that a horizontal $\bm{m}$-string and $\bm{f}$-string in the bulk are equivalent, up to enforced stabilizers, to:
\begin{align*}
\mathcal{L}_{\bm{m}} &\simeq \prod_{n=1}^{L} i\gamma_{n}\bar{\gamma}_n\equiv (-1)^F \\
\mathcal{L}_{\bm{f}} &= \mathcal{L}_{\bm{m}}\mathcal{L}_{\bm{e}} \simeq \prod_{n=1}^{L} i\gamma_{n}\bar{\gamma}_n \prod_{n=1}^{L-1} i\bar{\gamma}_n\gamma_{n+1}  \\
&= (-1) X_LX_{L+1/2}X_{1}i\bar{\gamma}_1\gamma_{L}.
\end{align*}
where $(-1)^F$ is the total fermion parity and $ X_LX_{L+1/2}X_{1}$ is the omitted boundary plaquette stabilizer on the original rough boundary. The bulk operator $\mathcal{L}_{\bm{f}}$ is therefore equivalent to the omitted boundary stabilizer on the fermionic boundary (up to a minus sign), which means that its eigenvalue corresponds to the boundary condition of the fermionic sandwich, which can be changed by inserting an $\bm{m}$-string vertically. The four sectors in the sandwich (labeled by total fermion parity and boundary condition) correspond to the eigenvalues of $\mathcal{L}_{\bm{m}}$ ($+1/ {-1}$ for even/odd parity) and $\mathcal{L}_{\bm{f}}$ ($+1/ {-1}$ for anti-periodic/periodic boundary condition).

\begin{figure}
    \centering
    \includegraphics[scale=0.55,page=69]{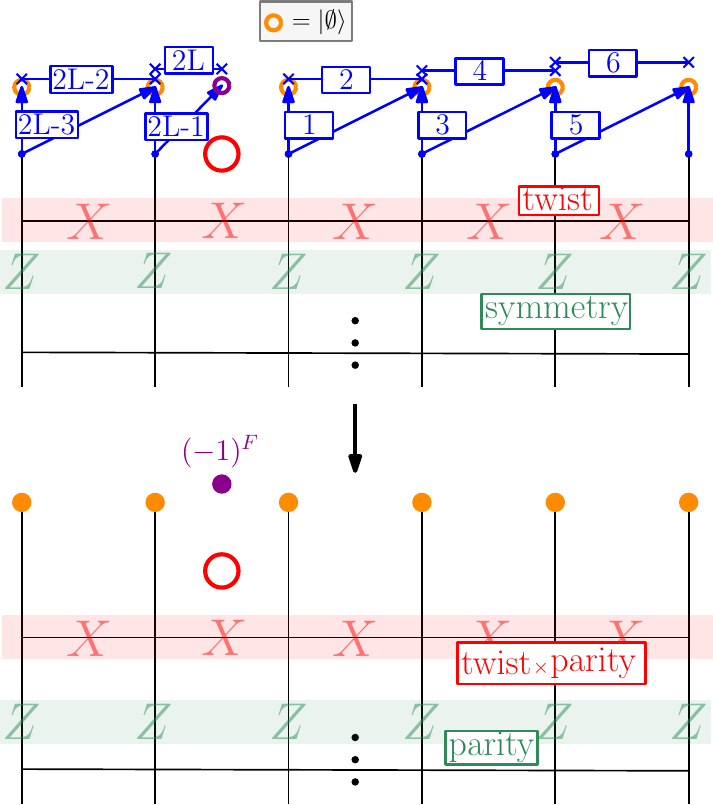}
    \caption{The circuit that maps a rough boundary to a fermionic boundary. The fermions are initialized on the empty state $\ket{\emptyset}$. The individual steps of the circuit are given in Eqs.~\ref{control_gate} and \ref{swap}. Every step of the circuit conserves the total fermion parity. After every step of the circuit, the occupation number of the fermion on the right corresponds to the total fermion parity of all the fermions on the left of the circuit. The circuit generates all the correct Hamiltonian terms of the fermionic boundary (Eq.~\ref{rough-fermionic}). In the last step of the circuit, an ancilla fermion is introduced (purple).}
    \label{fig:FermionicQC}
\end{figure}

We can write down a circuit that changes a rough boundary to the fermionic boundary. The circuit is shown in Fig.~\ref{fig:FermionicQC}. The fermionic degrees of freedom are initialized on the empty state ($\ket{\emptyset}$) (which is stabilized by $i\gamma_n\bar{\gamma}_n$). The gate that is controlled by every rough leg, that consists of two blue arrows, is a controlled$-i(\bar{\gamma}_n\bar{\gamma}_{n+1})$ gate:
\begin{align}
\label{control_gate}
\includegraphics[valign=c,page=71,scale=0.7]{duality_figures} = \frac{1}{2}(1+Z_{c}) + i\bar{\gamma}_n\bar{\gamma}_{n+1}\frac{1}{2}(1-Z_{c}),
\end{align}
where $Z_{c}$ acts on the control qubit of the rough leg. The fermionic swaps is defined as:
\begin{align}
\label{swap}
\includegraphics[valign=c,page=72,scale=0.7]{duality_figures} = \frac{1}{2}(1+\gamma_{n}\gamma_{n+1})(1+\bar{\gamma}_{n}\bar{\gamma}_{n+1}).
\end{align}
To ensure that fermion parity is preserved, each rough leg controls two majorana operators. Preserving fermion parity also necessitates the introduction of a single ancillary fermion in order to accommodate all parity sectors of the sandwich. The ancillary fermion has the same parity as the boundary fermions such that the total fermion parity of boundary + ancillary fermion is even. The correct boundary edge and plaquette stabilizers are recovered by the circuit as follows (except the one omitted plaquette stabilizer):
\begin{align} \label{rough-fermionic}
\begin{split}
\includegraphics[valign=c,page=67,scale=0.5]{duality_figures}\\
\includegraphics[valign=c,page=68,scale=0.5]{duality_figures}.
\end{split}
\end{align}
The last step of the circuit involves the ancilla fermion, in order to obtain the correct edge stabilizer on site $L$ while preserving the fermion parity in every step of the circuit. The occupation number of the ancilla fermion after the circuit corresponds to the total fermion parity of the boundary $i\gamma_a\bar{\gamma}_a = (-1)^F$ (where $a$ denotes the ancilla), since it is connected to a horizontal string of $Z$. Assuming the sandwich started in a definite symmetry sector (fixed eigenvalue of $\mathcal{L}_{\bm{m}}$), the ancillary fermion will be disentangled from the sandwich at the end of the circuit. More generally, we can perform a measurement on the ancilla fermion at the end to fix the total fermion parity.

We can now see how the bulk sector reinterpretation under the boundary change works. Since the twist is given by the eigenvalue of $\mathcal{L}_{\bm{f}}$, the eigenvalue of $\mathcal{L}_{\bm{e}}$ (which corresponds to the twist with the original rough boundary) corresponds to the twist (up to a minus sign) times the total parity. There are two gapped phases for fermions without any additional symmetry beyond $\mathbb{Z}_2^f$: a trivial superconductor and a topological superconductor (Kitaev chain) \cite{kitaev2001unpaired}. The two gapped phases can be obtained by fixing the top boundary on the fermionic boundary defined above and fixing the bottom boundary on the normal rough or smooth boundary. One can toggle between the two phases by either changing the bottom boundary or by changing the top boundary by applying the circuit that maps from a rough to a smooth boundary in Fig. \ref{KWseqCirc} to the fermionic boundary (the circuit does not touch the fermions when applied to the fermionic boundary). The gapped phases associated to the $2+1D$ toric code sandwich are summarized in Table \ref{tab:2dtc_phases}, where we allow for rough/smooth/fermionic on the top boundary and rough/smooth on the bottom boundary.

\begin{table}[]
    \centering
    \begin{tabular}{|c||ccc|}
    \hline
          &Rough&Smooth&Fermionic\\
          \hline\hline
        Rough & SSB & $\mathbb{Z}_2$ Sym. & Trivial SC \\
        Smooth & $\mathbb{Z}_2$ Sym. & SSB & Kitaev chain\\
        \hline
    \end{tabular}
    \caption{All $1+1D$ phases corresponding to different choices of the top boundary (columns)  or bottom boundary (rows) of the $2+1D$ toric code in the TH picture. We are only allowing for a fermionic boundary on top. Dualities correspond to permuting columns of the table. }
    \label{tab:2dtc_phases}
\end{table}

\section{String-nets and Dualities involving noninvertible symmetries}
\label{sec:SN}

In this section, we present the general formalism for dualities involving non-invertible (categorical) symmetries and the sequential quantum circuits that implement them in the $1+1D$ sandwich picture, where the bulk is a string-net model. We demonstrate the formalism by applying it to the $S_3$ quantum double, as it is the simplest model with non-abelian anyons that showcases duality beyond the strict definition in the TH formalism where the change in the top boundary condition can be induced by a symmetry of the bulk theory. This is because not all gapped boundaries---classified by the Lagrangian subalgebras of the bulk theory---are related by an anyon permutation that leaves the bulk invariant. We will omit several details about the construction and focus our discussion around the crucial differences for dualities involving categorical symmetries compared to the (simpler) abelian case. For an in-depth discussion of dualities with categorical symmetries and their explicit realization in $1+1D$, we refer the reader to Refs.~\onlinecite{Lootens2023, Lootens2024dualities, Lootens2023lowdepth}. Details surrounding the TH construction for non-abelian theories---and $S_3$ specifically---can be found in Refs.~\onlinecite{lin2023asymptotic, Huang2023topological}. 

\begin{figure}
    \centering
    \includegraphics[scale=0.65,page=73]{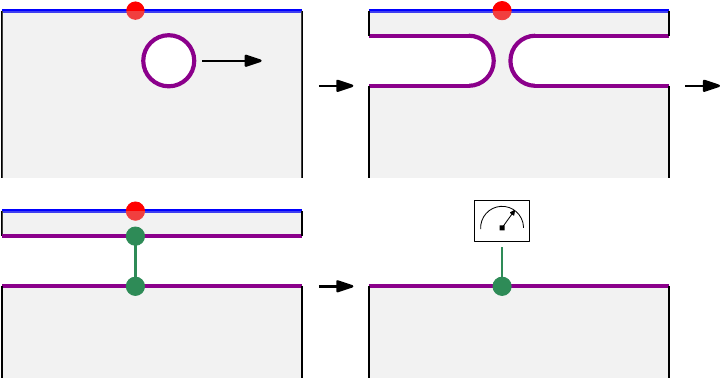}
    \caption{Schematic visualization of the procedure to change the gapped reference boundary of the bulk string-net from blue to purple. The red and green dot indicate possible boundary excitations on the old and new boundary respectively. First, a small bubble of the new boundary (purple) is opened up in the bulk, near the original boundary. As will be shown later, this step requires ancilla degrees of freedom. This bubble is then enlarged along the periodic direction of the sandwich using a SQC until the entire row is opened up, resulting in a bulk string-net with the new boundary that is connected to a sliver of string-net on top. The sliver can be further trivialized and the ancillary degrees of freedom can be measured out.}
    \label{fig:SNduality}
\end{figure}

The procedure for changing the reference boundary from one gapped string-net boundary to another, which corresponds to duality, is schematically sketched in Fig \ref{fig:SNduality} and explained in more detail later. The effect of the boundary change on the bulk topological sectors can be summarized as follows. In the presence of the gapped reference top boundary (blue), where we are allowing for boundary excitations to obtain all twist sectors (red dot), the bulk sectors (corresponding to the groundstates on a $2D$ torus of the bulk topological order) split into blocks that can be called boundary sectors. For abelian models (like the toric code), these boundary sectors are in one-to-one correspondence to the bulk sectors and each sector corresponds to a unique twist in the $1+1D$ model (given by the reference boundary excitation in the sandwich). For general non-abelian theories, this is not the case, because the non-abelian anyons can split according to the anyon condensation/confinement on the boundary in question. When we now change the boundary to study duality, creating excitations on the new boundary (green dot), the number of boundary sectors of the dual sandwich may not match the number of boundary sectors of the initial theory. However, the fundamental building blocks of the spectrum (called towers in the context of CFT) are still labeled by the bulk sectors, but they appear in the Hilbert space associated to different twists (labeled by the symmetry of the dual model) with possibly different exact degeneracies. In the explicit lattice construction of the $1+1D$ models in Refs.~\onlinecite{Lootens2023, Lootens2024dualities}, the symmetry acts on the twisted Hilbert space as elements of the so-called `tube algebra' \cite{ocneanu1994chirality, williamson2017symmetry, aasen2020topological} that correspond to the boundary sectors after a basis change. The boundary sectors further organize into `central idempotents’, which are labeled by elements in the Drinfeld center (or quantum double), corresponding to the bulk sectors \cite{evans1995ocneanu}. At the heart of dualities with general categorical symmetries lies therefore the mathematical equivalence known as `Morita equivalence' between the different categorical symmetries of the models dual to one another, meaning that their Drinfeld centers are equivalent as Modular Tensor Categories (MTC) \cite{muger2003subfactorsI,muger2003subfactorsII}, or in the TH setup, that the boundary sectors for different gapped boundaries organize into the same bulk sectors. Now we proceed by showing the general TH construction in $1+1D$ and the sequential quantum circuit that implements the duality, emphasizing the sector identification.

\subsection{String-nets in the TH picture}
\label{SNA}

The Hilbert space of a string-net model \cite{levin2005string} consists of configurations on a hexagonal lattice, with the edges labeled by simple objects of a unitary fusion category $\mathcal{C}$ (called the input category). The string-net model will act as the $2+1D$ bulk of the sandwich construction. The string-net Hamiltonian is a sum of commuting projectors
\begin{equation} \label{SNhamiltonian}
    H_{\text{SN}}^{\text{bulk}} = -\sum_{v} A_v - \sum_{p} B_p.
\end{equation}
The vertex terms ($A_v$) enforce the fusion rules of the simple objects in $\mathcal{C}$ at every vertex and the plaquette terms $(B_p)$ give dynamics to the string-net: $B_p = \sum_s (d_s/D) B_p^s$, with $d_s$ the quantum dimensions of the simple objects $s$, $D = \sum_s (d_s)^2$ the total quantum dimension and $B_p^s$ the action of fusing a single loop $s$ in the lattice as defined in Ref.~\onlinecite{levin2005string}. The bulk anyons (denoted by greek letters in the following discussion) are given by simple objects in the Drinfeld center of $\mathcal{C}$ $\alpha \in Z(\mathcal{C})$ (a Unitary Modular Tensor Category, or UMTC).

The possible gapped boundaries are classified by the Lagrangian subalgebras $\mathcal{A} = \sum_{\alpha} w_{\alpha} \alpha$, with $w_{\alpha}$ non-negative integers, of the bulk UMTC. The anyons where $w_{\alpha}>0$ are the anyons that condense on the boundary \cite{cong2016topological, cong2017hamiltonian}. For non-abelian anyons with $w_{\alpha}>1$ there are multiple condensation channels. Alternatively, a gapped boundary is given by an indecomposable (left) module category $\mathcal{M}$ of $\mathcal{C}$ \cite{kitaev2012models}, due to the 1-1 correspondence between module categories of $\mathcal{C}$ and Lagrangian subalgebras of $Z(\mathcal{C})$ \cite{davydov2013witt}. The degrees of freedom on the boundary are labeled by simple objects in $\mathcal{M}$. Boundary excitations are labeled by the confined anyons after condensation (after possible identification and splitting), which can be labeled by the simple objects in the fusion category $\mathcal{C}_{\mathcal{M}}^*$ (the Morita dual of $\mathcal{C}$ with respect to $\mathcal{M}$), which acts as the (categorical) symmetry of the sandwich. The gapped boundary $\mathcal{M}=\mathcal{C}$, sometimes called the canonical boundary, can be seen as the generalization of the smooth boundary in the toric code. In this case, the symmetry of the sandwich $\mathcal{C}_{\mathcal{M}}^*$ is simply $\mathcal{C}$ itself. When the input category $\mathcal{C}$ is modular and $Z(\mathcal{C}) = \mathcal{C} \boxtimes \overline{\mathcal{C}}$, this boundary is obtained by the Lagrangian algebra $\mathcal{A} = \sum_{a \in \mathcal{C}}(a,\overline{a})$ (condensation of the charge excitations) \footnote{Note that on the canonical boundary, the charge excitations condense. In the toric code discussion, it was said that the flux excitations condense on the smooth (canonical) boundary, in order to be consistent with our definition of the toric code Hamiltonian.}. \\

\begin{figure}
    \includegraphics[width=\linewidth,scale=1,page=62]{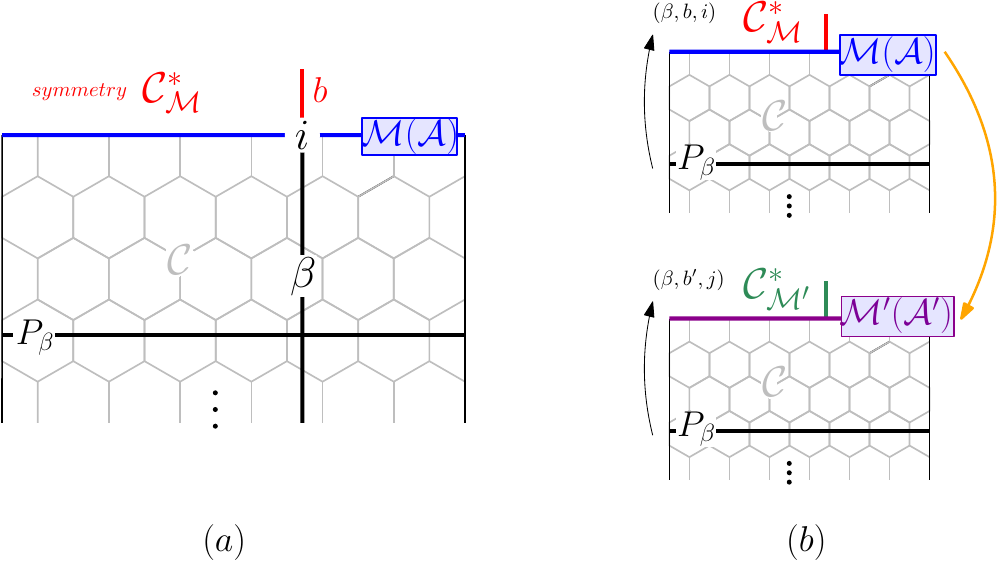}
    \caption{(a) The sandwich (TH) picture for general string-nets with input category $\mathcal{C}$ and top boundary $\mathcal{M}$ (condensable Lagrangian algebra $\mathcal{A}$). The corresponding symmetry is given by the fusion category $\mathcal{C}_{\mathcal{M}}^*$. Allowing for all boundary excitations, the total Hilbert space of the sandwich can be decomposed into bulk sectors $\beta \in Z(\mathcal{C})$ through the projectors in Eq.~\ref{HilbertSpaceBeta}, which act in the horizontal direction below the boundary. The bulk sectors decompose further into boundary sectors $(\beta,b,i)$, labeled by the boundary excitation $b \in \mathcal{C}_{\mathcal{M}}^*$ and the confinement channel $i$ of the corresponding anyon that tunnels to the top boundary. (b) Duality corresponds to changing the top boundary from $\mathcal{M}(\mathcal{A})$ to $\mathcal{M'}(\mathcal{A'})$. The symmetry of the new sandwich is $\mathcal{C}_{\mathcal{M'}}^*$. The same bulk sector $\beta$ decomposes differently on the boundary of the dual sandwich.}
    \label{general_sandwich}
\end{figure}

The sandwich setup is shown in Fig.~\ref{general_sandwich}. The top boundary is fixed on $\mathcal{M}$ (condensing anyons in the subalgebra $\mathcal{A}$). The symmetry of the effective $1+1D$ model is determined by this choice of boundary and given by $\mathcal{C}_{\mathcal{M}}^*$, as are the excitations on the boundary. We consider the full sandwich Hilbert space, where we allow for all types of boundary excitations at one boundary defect point. In the toric code sandwich, this was obtained by not enforcing one single boundary stabilizer. For general string-nets, this is obtained by adding an excitation (red line in Fig.~\ref{general_sandwich}), labeled by elements in $\mathcal{C}_{\mathcal{M}}^*$, on the boundary. This corresponds to allowing for all twisted boundary conditions in the corresponding $1+1D$ Hilbert space. The total Hilbert space of the sandwich can be decomposed according to the bulk sectors (as though the system was defined on a torus) in the Minimal Entanglement States (MES) basis introduced in Ref.~\cite{zhang2012quasiparticle}. In this basis, the torus groundstates are eigenstates of the Wilson lines in the horizontal direction (along the boundary), denoted by $\mathcal{L}_{\alpha}$. The sectors are obtained by applying horizontal projectors $P_{\beta}$, defined with the help of the $S$-matrix: 
\begin{align} 
\label{HilbertSpaceBeta}
\mathcal{H}^{\mathcal{M}(\mathcal{A})} &= \bigoplus_{\beta} \
\includegraphics[valign=c,page=44,scale=0.6]{duality_figures}, \ \\
P_{\beta} &= \sum_{\gamma} S_{1\beta} \bar{S}_{\beta\gamma} \ \mathcal{L}_{\gamma} .
\label{Projectors}
\end{align}

The advantage of this basis is that it is defined independent of the top boundary and is therefore unaffected when the boundary is changed through a sequential quantum circuit (see next section). The ground states on the torus in the MES basis can be obtained by inserting an anyon $\beta$ in the vertical direction, starting from the ground state defined by $P_1$. Going to the cylinder geometry, the sectors decompose on the boundary according to the behaviour of the vertically inserted anyon on the boundary. For Abelian theories, anyons are confined on the boundary in a simple boundary excitation, meaning that every bulk sector has a definite twist in the $1+1D$ Hamiltonian. In the non-abelian case, confined anyons may not correspond to simple boundary excitations. Rather, when we insert an anyon in the vertical direction, it can split on the boundary in multiple condensation/confinement channels: $\beta = \sum_{b} W_{\beta}^b \ b$, where $W_{\beta}^b$ is a positive integer that characterizes the multiplicity of the bulk anyon $\beta$ decomposing into $b$ \cite{bais2002broken, bais2009condensate, shen2019defect}. Note that for $b=1$, $W_{\beta}^1 = w_{\beta}$. The Hilbert space further decomposes as 
\begin{equation}
\label{boundarySectors}
\mathcal{H}^{\mathcal{M}(\mathcal{A})}_{\beta} = \bigoplus_{b} W_{\beta}^b \ \mathcal{H}_{\beta}^{b, \mathcal{M}(\mathcal{A})}, 
\end{equation}
and the boundary sectors are now labeled by a triplet $\beta \to (\beta,b,i)$ where $b$ is chosen such that $W_{\beta}^b>0$ and $i=1,\dots,W_{\beta}^b$. The sector decomposition in the MES basis stems from the fact that the bulk Wilson lines in Eq.~\ref{HilbertSpaceBeta}, when brought close to the boundary, interact non-trivially with it. In the process of wrapping an anyon around the cylinder, close to the boundary, it can braid non-trivially with a condensed anyon that is created from the boundary, a process that is characterized through a topological number called the half-braiding (or half-linking) \cite{bais2002broken, bais2009condensate, shen2019defect}. In appendix \ref{S3data} we show for the $S_3$ quantum double how the bulk Wilson lines decompose differently on different boundaries relevant for the duality studied in section \ref{sec:2DS3} and how as a result the bulk sectors in Eq. \ref{boundarySectors} decompose into different boundary sectors for the two sandwiches. 

Since duality corresponds to changing the top boundary in the TH formalism from $\mathcal{M}$ (condensable algebra $\mathcal{A}$) to $\mathcal{M}'$ (condensable algebra $\mathcal{A}'$), the bulk sectors, defined in Eq.~\ref{HilbertSpaceBeta} through bulk Wilson lines, are not affected by this change. Only the way the bulk sectors decompose on the boundary changes. This is consistent with the fact that duality induces a one-to-one correspondence between the bulk topological sectors of both theories. Generally, the numbers of boundary sectors of both theories don't match, only the bulk sectors are in one-to-one correspondence. This means that a sector in the low-energy spectrum of the theories dual to one another can come with different degeneracies. A sector $\beta$ appears in the initial theory in the Hamiltonian with twist $b \in \mathcal{C}_{\mathcal{M}}^*$ with degeneracy $W_{\beta}^b$. The same sector appears in the dual theory in the Hamiltonian with twist $b' \in \mathcal{C}_{\mathcal{M}'}^*$ with degeneracy $W_{\beta}^{b'}$. It is therefore more accurate for non-abelian theories to state that duality preserves the spectrum `up to exact degeneracies'.  We will illustrate the correspondence between bulk and boundary sectors for the $S_3$ quantum double in section \ref{sec:2DS3}.

\subsection{Sequential quantum circuits for string-net dualities}
\label{sec:SNSQC}

Now, we are ready to present the sequential quantum circuit that changes the top boundary in the sandwich setup outlined above. A low-depth circuit for general $1+1D$ dualities with categorical symmetries was given in Ref.~\onlinecite{Lootens2023lowdepth} and it was shown explicitly how the circuit requires ancilla degrees of freedom in order to be unitary. In light of the discussion in the previous section, it is clear that the ancilla degrees of freedom are required because the boundary sectors in the initial and dual theory associated to the same bulk sector $\beta$ may not match. The ancilla degrees of freedom enlarge the Hilbert space, taking account of the exact degeneracies in the spectrum associated to this mismatch. The degeneracies can be lifted by performing a measurement on the ancilla degrees of freedom.

\begin{figure}
\centering
    \includegraphics[width=\linewidth,scale=0.8,page=64]{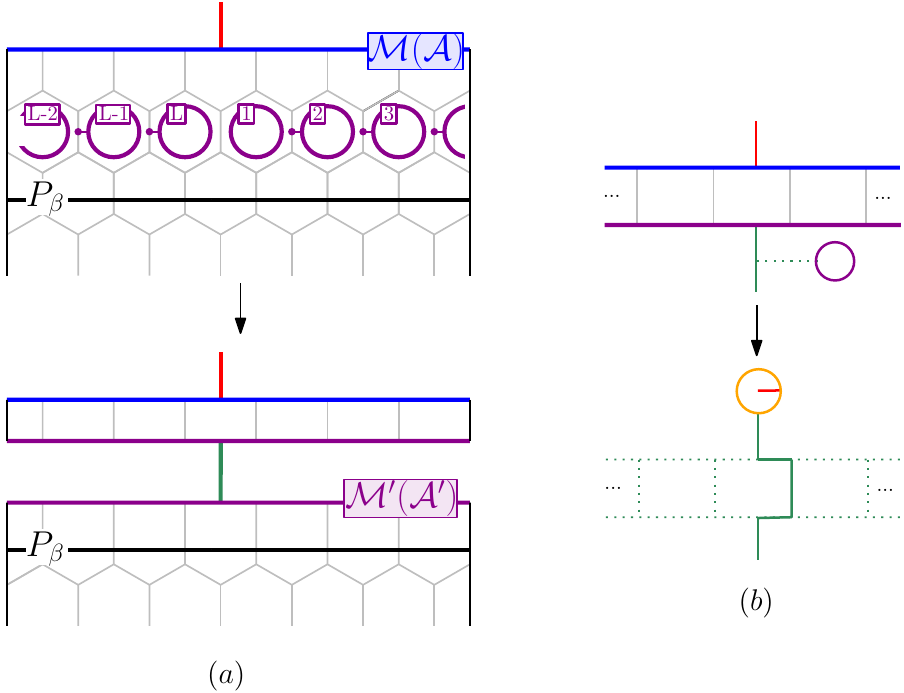}
    \caption{(a) The sequential quantum circuit (purple plaquette operators) to map from a boundary $\mathcal{M}(\mathcal{A})$ (in blue), to a boundary $\mathcal{M}'(\mathcal{A}')$ (in purple) by opening up a row of the new boundary, slightly below the old boundary. The individual steps in the circuit are given by Eqs. \ref{firstStep}-\ref{LastStep}. A bulk sector $\beta$ is selected by using the projectors defined in Eq.~\ref{HilbertSpaceBeta}. The circuit acts as an isometry, where the Hilbert space is enlarged with ancilla degrees of freedom, shown by the boxes in Eqs.~\ref{firstStep}-\ref{LastStep}. (b) Using these ancilla qudits, the remaining strip of string-net can be trivialized through a second circuit that is diagramatically similar to the first one but requires more complicated recouplings. The ancilla qudits after this step are in one-to one correspondence to the ones produced by the circuit in Ref.~\cite{Lootens2023lowdepth}.}
    \label{string-net_circuit}
\end{figure}

Here, we present a circuit, in the spirit of the sequential quantum circuits used in Ref.~\onlinecite{Chen2024}, that, acts on one row in the bulk, slightly below the boundary. The circuit, and its effect on the sandwich, is shown in Fig.~\ref{string-net_circuit}. The circuit establishes the same duality mapping and sector identification as the one in Ref.~\cite{Lootens2023lowdepth}. The circuit works by essentially cutting off the old boundary and replacing it with a new one.
We allow for all boundary excitation types at one defect point on the boundary and fix the initial bulk sector on $\beta$ by the horizontal projection $P_{\beta}$ (Eq.~\ref{HilbertSpaceBeta}). It is clear from the previous discussion that the same bulk sector $\beta$ can appear in the spectrum of the dual theory with different exact degeneracies (this will be illustrated for the $S_3$ quantum double in the next section). Therefore, we don't expect the circuit to act as a unitary, but as an isometry. The circuit acts on an enlarged Hilbert space, where one has to introduce ancilla qudits, to take into account this possible mismatch in the boundary sectors of both theories that correspond to the same bulk sector $\beta$.

The sequential quantum circuit has a depth that scales linearly in the system size in the periodic direction and acts by applying a series of plaquette operators (purple in Fig.~\ref{string-net_circuit}) labeled by the module category of the target boundary $\mathcal{M}'$ \footnote{Technically, the circuit relies on the structure of $\mathcal{M}'$ as a $(\mathcal{C},\mathcal{C}_{\mathcal{M}'}^*)$-bimodule, see Ref.~\onlinecite{lootens2022mapping}, where, given $\mathcal{M}'$ as a left-module category of $\mathcal{C}$, $\mathcal{C}_{\mathcal{M}'}^*$ is the unique Morita equivalent category of $\mathcal{C}$ for which $\mathcal{M}'$ is also a right module category.}. We will diagrammatically sketch the action of the circuit on the string-net state. The details of the individual steps in terms of the $F$-symbols can be found in Ref.~\onlinecite{Chen2024}. The unitarity of the circuit relies on the unitarity of the $F$-symbols involved in the individual steps of the circuit. In the following steps of the circuit we will represent the ancilla qudits with squares and the lattice qudits that are changed under the action of the circuit with circles. The color indicates the category whose simple objects label the qudit (grey for $\mathcal{C}$, purple for $\mathcal{M}'$ and green for $\mathcal{C}_{\mathcal{M}'}^*$). 

In the first step of the circuit, a bubble of  $\mathcal{M}'$ (the module category for the target boundary) is opened on the first plaquette. This step requires enlarging the local Hilbert space with three ancilla degrees of freedom:
\begin{align} \label{firstStep}
\includegraphics[valign=c,page=16,scale=0.85]{duality_figures}.
\end{align}
One ancilla qudit (purple box) is initialized on the superposition $\sum_{S\in \mathcal{M}'} \frac{d_S}{D^2} \ket{S}$ (with $d_S$ the quantum dimension of the simple object $S\in \mathcal{M}'$), another qudit is initialized on the trivial element in $\mathcal{C}$ (empty grey square) and the third ancilla qudit is a copy of one of the lattice qudits (filled grey square). Next, a series of (unitary) $F$-moves is applied to fuse a purple loop into the grey lattice controlled by the purple ancilla. 
The next steps in the circuit are controlled plaquette operators acting on the plaquettes $2,\dots,L-1$ and are the inverse operators of the ones in Ref.~\onlinecite{Chen2024} used to construct a string-net state from a product state:
\begin{align} \label{nextSteps}
\includegraphics[valign=c,page=17,scale=0.85]{duality_figures}.
\end{align}
The step consists of fusing the purple control qudit on the left edge of the plaquette with the grey qudits around the plaquette, moving the purple line one plaquette further. At the end of this step, the control qudit is mapped to the trivial object in $\mathcal{C}_{\mathcal{M}'}^*$ (empty green circle). As a result, the two purple qudits connected to the trivial green line both on the top and on the bottom are identified and they represent the new boundary qudits. The step is repeated along the entire row. In the last plaquette operation of the circuit ($L$), the last and first plaquette become: 
\begin{align} \label{LastStep}
\includegraphics[valign=c,page=18,scale=0.85]{duality_figures}.
\end{align}
The remaining top row is entangled with the bulk only through the green line, which labels the excitation on the new boundary, and the ancilla purple qudit.

After this circuit, we have a new boundary labeled by $\mathcal{M}'$ that is connected by a single edge to a small strip of string-net on the top and a collection of ancilla qudits that are still entangled with the bulk. We can trivialize the remaining strip of string-net on top by applying a circuit that is diagramatically similar to the circuit that opens up the row (descirbed above), but requires more complicated recouplings that fuse purple and blue lines into orange lines ---labeled by objects in the category of module functors $\text{Fun}_{\mathcal{C}}(\mathcal{M},\mathcal{M}')$---. However, we do not wish to go into further detail about this step and the related mathematical structure, but simply sketch the effect of this step on the ancilla qudits in Fig.\ref{string-net_circuit} (b) \footnote{When we trivialize the remaining strip on top, the collection of ancilla qudits that are still entangled with the bulk now effectively label what are known as `intertwiner tubes'. After a suitable unitary on these degrees of freedom, they label the boundary sectors of the dual model in the same way as in ~\cite{Lootens2023lowdepth}.}.
Because we have fixed the bulk sector from the start on $\beta$, only the boundary sectors of the dual model $(\beta,b',j)$ that are consistent with the bulk sector give non-vanishing contributions to the spectrum. The circuit preserves the spectrum, but can change the exact degeneracy. One can lift these degeneracies by performing a measurement on the ancilla qudits. Performing a measurement on the new boundary excitation ($b'$) is not sufficient to entirely lift the degeneracy, since there is a further degeneracy when  $W_{\beta}^b>1$. Note that in our diagrams, we have omitted any degeneracy labels in the fusion spaces between the objects for visual simplicity. We have to take these into account to fully lift the degeneracies in the spectrum associated to $W_{\beta}^b>1$.

Finally, we note that the Kramers-Wannier duality (toric code sandwich) can be recovered by the following choices for the categories involved: $\mathcal{C} = \text{Vec}_{\mathbb{Z}_2}$,
$\mathcal{M} = \text{Vec}_{\mathbb{Z}_2}$,  $\mathcal{M'} =  \text{Vec}$, $\mathcal{C}_{\mathcal{M}}^* = \text{Vec}_{\mathbb{Z}_2}$ and $\mathcal{C}_{\mathcal{M}'} = \text{Rep}(\mathbb{Z}_2)$. The mapping $\mathcal{M}\rightarrow \mathcal{M}'$ corresponds to the duality from a smooth to a rough boundary. The correspondence between bulk and boundary sectors in the toric code is one-to-one and therefore we don't require ancilla qubits.

\subsection{$S_3$ Quantum Double}
\label{sec:2DS3}

\begin{table*}
\caption{The behaviour of the bulk sectors $\beta$ (labeled by a conjugacy class of $S_3$ and an irrep of the centralizer) on all four gapped boundaries of the $S_3$ quantum double, characterized by a condensable Lagrangian algebra $\mathcal{A}$ or by the corresponding module category $\mathcal{M}$. For every boundary, the corresponding symmetry of the sandwich ($\mathcal{C}_{\mathcal{M}^*}$) is shown and the bulk anyons either fully condense (cond.) or split into multiple simple objects in $\mathcal{C}_{\mathcal{M}^*}$. In the $\text{Vec}_{S_3}$ case, the different boundary sectors are labeled by the six group elements of $S_3$ and in the Rep$(S_3)$ case, by the simple objects $1,\rho_-,\rho_2$.}
\label{anyonsS3}
\begin{tabular}{|ccccccccccc|} 
\hline
    Bulk sector $\beta$ & && $\bm{A}$  & $\bm{B}$  & $\bm{C}$ & $\bm{D}$ & $\bm{E}$ & $\bm{F}$ & $\bm{G}$ & $\bm{H}$\\
& && $[1,1]$  &$[1,\rho_-]$  & $[1,\rho_2]$ & $[s,+]$ & $[s,-]$ & $[r,1]$ & $[r,\omega]$& $[r,\bar{\omega}]$\\
\hline
cond. algebra&$\mathcal{M}$&Symmetry&   &  &  &  &  &  & &\\
\hline
$\mathcal{A}_1 = \bm{A} + \bm{B} + 2\bm{C}$& $\text{Vec}_{S_3}$ & $\text{Vec}_{S_3}$ & cond. & cond.  & \begin{tabular}{@{}c@{}}cond. \\ $1,1$\end{tabular} & \begin{tabular}{@{}c@{}}split \\ $s,sr,s\overline{r}$\end{tabular} & \begin{tabular}{@{}c@{}}split \\ $s,sr,s\overline{r}$\end{tabular} & \begin{tabular}{@{}c@{}}split \\ $r,\overline{r}$\end{tabular} & \begin{tabular}{@{}c@{}}split \\ $r,\overline{r}$\end{tabular} & \begin{tabular}{@{}c@{}}split \\ $r,\overline{r}$\end{tabular}\\
\hline
$\mathcal{A}_2 = \bm{A} + \bm{B} + 2\bm{F}$& $\text{Vec}_{\mathbb{Z}_2}$ & $\text{Vec}_{S_3}$ & cond. & cond.  & \begin{tabular}{@{}c@{}}split \\ $r,\overline{r}$\end{tabular} & \begin{tabular}{@{}c@{}}split \\ $s,sr,s\overline{r}$\end{tabular} & \begin{tabular}{@{}c@{}}split \\ $s,sr,s\overline{r}$\end{tabular} & \begin{tabular}{@{}c@{}}cond. \\ $1,1$\end{tabular} & \begin{tabular}{@{}c@{}}split \\ $r,\overline{r}$\end{tabular} & \begin{tabular}{@{}c@{}}split \\ $r,\overline{r}$\end{tabular}\\
\hline
$\mathcal{A}_3 = \bm{A} + \bm{C} + \bm{D}$& $\text{Vec}_{\mathbb{Z}_3}$ & $\text{Rep}(S_3)$ & cond. & $\rho_-$  & \begin{tabular}{@{}c@{}}split \\ $1,\rho_-$\end{tabular} & \begin{tabular}{@{}c@{}}split \\ $1,\rho_2$\end{tabular} & \begin{tabular}{@{}c@{}}split \\ $\rho_-,\rho_2$\end{tabular} & $\rho_2$ & $\rho_2$ & $\rho_2$\\
\hline
$\mathcal{A}_4 = \bm{A} + \bm{F} + \bm{D}$& $\text{Vec}$ & $\text{Rep}(S_3)$ & cond & $\rho_-$ & $\rho_2$ & \begin{tabular}{@{}c@{}}split \\ $1,\rho_2$\end{tabular} & \begin{tabular}{@{}c@{}}split \\ $\rho_-, \rho_2$\end{tabular}  & \begin{tabular}{@{}c@{}}split \\ $1,\rho_-$\end{tabular} & $\rho_2$ & $\rho_2$\\
\hline
\end{tabular}
\end{table*}

The $S_3$ quantum double is a model with non-abelian anyons that showcases duality in the TH formalism beyond the strict definition where a symmetry of the bulk is applied. In this section, we focus on the dualities that arise in our sandwich setup where the bulk is the $S_3$ quantum double. Due to the non-abelian nature of the bulk theory, some of the dualities involve the non-invertible (categorical) symmetry Rep$(S_3)$. We illustrate, in the most direct way, how the bulk sectors decompose differently on the gapped boundaries through the condensation/confinement behaviour of the vertical anyon $\beta$ in Fig. \ref{general_sandwich}. In Appendix \ref{S3data}, we show how this decomposition works explicitely in the MES basis of Eq. \ref{HilbertSpaceBeta} by decomposing the horizontal bulk Wilson lines $\mathcal{L}_{\alpha}$ on the boundary.

The input category for the $S_3$ quantum double is $\text{Vec}_{S_3}$, the category of vector spaces over $S_3$. We label the six group elements of $S_3$ as ${1,r,\overline{r},s,sr,s\overline{r}}$ with $s^2=r^3=(sr)^2=1$. The bulk anyons are given by the simple objects in the Drinfeld center $\alpha \in Z(\text{Vec}_{S_3}) \simeq D(S_3)$. The $S_3$ quantum double has eight anyons (bulk sectors), labeled by $\bm{A},\bm{B},\bm{C},\dots,\bm{H}$: $\bm{A}$, $\bm{B}$, $\bm{C}$ are the trivial, sign and 2-dimensional irreps of $S_3$ respectively and are pure charge excitations, $\bm{D},\bm{F}$ are the pure flux excitations and $\bm{E},\bm{G},\bm{H}$ are dyons. $\bm{C},\bm{F},\bm{G},\bm{H}$ are 2-dimensional and $\bm{D}, \bm{E}$ are 3-dimensional. Of these anyons, $\bm{A},\bm{B},\bm{C},\bm{D}$ and $\bm{F}$ are bosons and can condense \cite{kitaev2003fault, beigi2011quantum}. 

For quantum doubles, the indecomposable module categories are given by subgroups $H$ of $G$ and elements in the second cohomology group $H^{2}(H,U(1))$ \cite{etingof2016tensor}. The possible boundaries for $D(S_3)$ are summarized in Table \ref{anyonsS3}. They are labeled by the Lagrangian subalgebra ($\mathcal{A}$) which in turn determines the corresponding module category ($\mathcal{M}$) and the symmetry of the corresponding $1+1D$ model ($\mathcal{C}_{\mathcal{M}}^*$). The behaviour of the anyons under the condensation is shown, i.e. how they split into elements of $\mathcal{C}_{\mathcal{M}}^*$ on the boundary. This splitting can be obtained by following the condensation procedure outlined in Ref.~\onlinecite{eliens2014diagrammatics}.
Regardless of the boundary, we fix the bulk topological sector of the sandwich by applying the corresponding projectors $P_{\beta}$ (Eq.~\ref{HilbertSpaceBeta}) in the horizontal direction, below the boundary. The eight projectors are given in Eq.~\ref{S3sectors}. The bulk sectors can be labeled by a conjugacy class of $S_3$, $C_1=\{1\}$, $C_r=\{r,\bar{r}\}$, $C_{s}=\{s,sr,s\bar{r}\}$, and an irrep of the centralizer. The pure charge sectors $\bm{A},\bm{B},\bm{C}$ have trivial conjugacy class $C_1$. The corresponding irreps are the irreps of $S_3$, labeled by $1,\rho_-,\rho_2$. The sectors $\bm{D}$ and $\bm{E}$ have conjugacy class $C_s$ and the centralizer is $\mathbb{Z}_2$. Finally, the sectors $\bm{F}$, $\bm{G}$ and $\bm{H}$ have conjugacy class $C_r$ and the centralizer is $\mathbb{Z}_3$.

The way the bulk sectors $\beta$ decompose into boundary sectors ($\beta,b,i$) can be read off from table \ref{anyonsS3}. The bulk sectors correspond to the eight columns labeled by the anyons. Each sector decomposes into different blocks according to the splitting of the corresponding anyon $\beta$ that is inserted vertically, on the boundary. In Appendix \ref{S3data}, we show how the bulk sectors decompose into the boundary sectors for both $\mathcal{A}_1$ and $\mathcal{A}_4$ by decomposing the horizontal Wilson lines $\mathcal{L}_{\alpha}$ in Eq. \ref{HilbertSpaceBeta} on the boundary (see Fig. \ref{general_sandwich}). On the canonical boundary $\mathcal{A}_1$ (the charge condensed boundary) for example, the excitations are labeled by elements of Vec$_{S_3}$. The anyons $\bm{A}, \bm{B}$ and $\bm{C}$ are condensed. $\bm{D}, \bm{E}$ are identified and split into 3 excitations and $\bm{F}$, $\bm{G}, \bm{H}$ are identified and split into 2 excitations. The bulk sectors $\bm{D},\dots,\bm{H}$ decompose according to the number of elements in the corresponding conjugacy class. The anyon $C$ has 2 condensation channels, so the corresponding sector splits into two blocks $(\bm{C},1,1)$ and $(\bm{C},1,2)$. On the canonical boundary, there are a total of 16 boundary sectors. The splitting of bulk sectors into boundary sectors  for the canonical boundary is given in Eq. \ref{sectorsplittingA1}.
On the boundary $\mathcal{A}_4$ (the flux condensed boundary), the excitations are labeled by elements of the categorical symmetry Rep$(S_3)$. The anyons $\bm{D}$ and $\bm{F}$ are partially condensed and split in two blocks and $\bm{E}$ is fully confined and also splits into two blocks. The anyons $\bm{G}, \bm{H}$ are fully confined and correspond to the excitation $\rho_2$. There are a total of 11 boundary sectors. The splitting of bulk sectors into boundary sectors for the flux-condensed boundary is given in Eq. \ref{boundarySectorsA4}.

The dualities $\mathcal{A}_1 \to \mathcal{A}_2$ and $\mathcal{A}_3 \to \mathcal{A}_4$ correspond to a symmetry of the bulk topological order in the TH formalism by exchanging the anyons $\bm{C}$ and $\bm{F}$. The duality $\mathcal{A}_1 \to \mathcal{A}_4$ is more exotic since it relates a $1+1D$ system with group symmetry $S_3$ to a system with categorical symmetry Rep$(S_3)$, with the non-trivial fusion rule $\rho_2 \times \rho_2 = 1 + \rho_- + \rho_2$. As a consequence, the Hilbert space with non-trivial twist $\rho_2$ contains the five bulk sectors $\bm{D},\bm{E},\bm{F},\bm{G}$ and $\bm{H}$ (see Eq. \ref{boundarySectorsA4}) while the trivial and $\rho_-$ sectors constain three sectors each. To illustrate the `mismatch' in the exact degeneracies in the spectrum between the two theories, we can take the bulk sector $\bm{C}$ in table \ref{tab:S3_phases} as an example. In the sandwich with a flux-condensed boundary ($\mathcal{A}_4$), this sector appears once in the $\rho_2$-twisted Hilbert space. In the sandwich with charge-condensed boundary ($\mathcal{A}_1$), this sector appears with exact two-fold degeneracy in the untwisted Hilbert space. Therefore, when we implement the duality $\mathcal{A}_4 \to \mathcal{A}_1$ and we fix the bulk sector on $\bm{C}$, the sequential quantum circuit will enlarge the Hilbert space and produce an exact degeneracy in the spectrum. One needs to perform a measurement on the ancilla qudits at the end of the circuit to lift this two-fold degeneracy.

We end this section by briefly showing the effect of duality on the gapped phases associated to the $S_3$ quantum double sandwich. Gapped phases are discussed in many other works, see for example Ref.~\onlinecite{Huang2023topological} for $S_3$ gapped phases in the TH formalism and Ref.~\onlinecite{garre2023classifying} for gapped phases in $1+1D$ with general categorical symmetries.

A gapped phase can be obtained by fixing  both the top and bottom boundary of the sandwich on a gapped boundary such that the different symmetry breaking patterns are recovered by the anyons that are allowed to condense on both boundaries. The symmetry of the sandwich is still dictated by the choice of top boundary $\mathcal{M}(\mathcal{A})$. 

For $1+1D$ systems with group symmetry $G$ or categorical symmetry Rep($G$), the gapped phases are given by subgroups $H$ of $G$ and the elements in the second cohomology group $H^2(H,U(1))$ \cite{etingof2016tensor}. For $S_3$, all subgroups have a trivial second cohomology group and the four gapped phases, labeled by spontaneous symmetry breaking (SSB) to all of the subgroups of $S_3$ ($1$,$\mathbb{Z}_2$,$\mathbb{Z}_3$ and $S_3$) are shown in table \ref{tab:S3_phases} in terms of the choice of top and bottom boundary. Given a choice of bottom boundary (row), duality corresponds to permuting the colummns of this table. This concludes our discussion on dualities in the TH formalism for general bulk string-nets.

\begin{table}[]
    \centering
    \begin{tabular}{|c||cccc|}
    \hline
          & \begin{tabular}{@{}c@{}} $\mathcal{A}_1$ \\ Vec($S_3$) \end{tabular} & \begin{tabular}{@{}c@{}} $\mathcal{A}_2$ \\ Vec($S_3$) \end{tabular} & \begin{tabular}{@{}c@{}} $\mathcal{A}_3$ \\ Rep($S_3$) \end{tabular} &  \begin{tabular}{@{}c@{}} $\mathcal{A}_4$ \\ Rep($S_3$) \end{tabular} \\\hline\hline
        $\mathcal{A}_1$ & SSB $\to$ 1 & SSB $\to$ $\mathbb{Z}_3$ & SSB $\to$ $\mathbb{Z}_2$ &Sym.\\
        $\mathcal{A}_2$ & SSB $\to$ $\mathbb{Z}_3$ & SSB $\to$ 1 & Sym. &SSB $\to$ $\mathbb{Z}_2$\\
        $\mathcal{A}_3$ & SSB $\to$ $\mathbb{Z}_2$ & Sym.& SSB $\to$ 1&SSB $\to$ $\mathbb{Z}_3$\\
        $\mathcal{A}_4$ & Sym. & SSB $\to$ $\mathbb{Z}_2$& SSB $\to$ $\mathbb{Z}_3$ &SSB $\to$ 1\\
        \hline
    \end{tabular}
    \caption{All $1+1D$ gapped phases corresponding to different choices of the top boundary (columns) or bottom boundary (rows) of the $S_3$ quantum double in the TH picture. Dualities correspond to permuting columns of the table. The fully symmetric phase is denoted by `Sym.' and `SSB $\to H$' means the partial symmetry broken phase with $H$  the remaining symmetry. The choice of top boundary affects whether the system is regarded as having a Vec$_{S_3}$ or a Rep$(S_3)$ symmetry, as indicated.}
    \label{tab:S3_phases}
\end{table}

\section{$3+1D$ Toric Code and higher form symmetries}
\label{sec:3DTC}

\begin{figure}[b]
    \centering
    \includegraphics[scale=0.65,page=58]{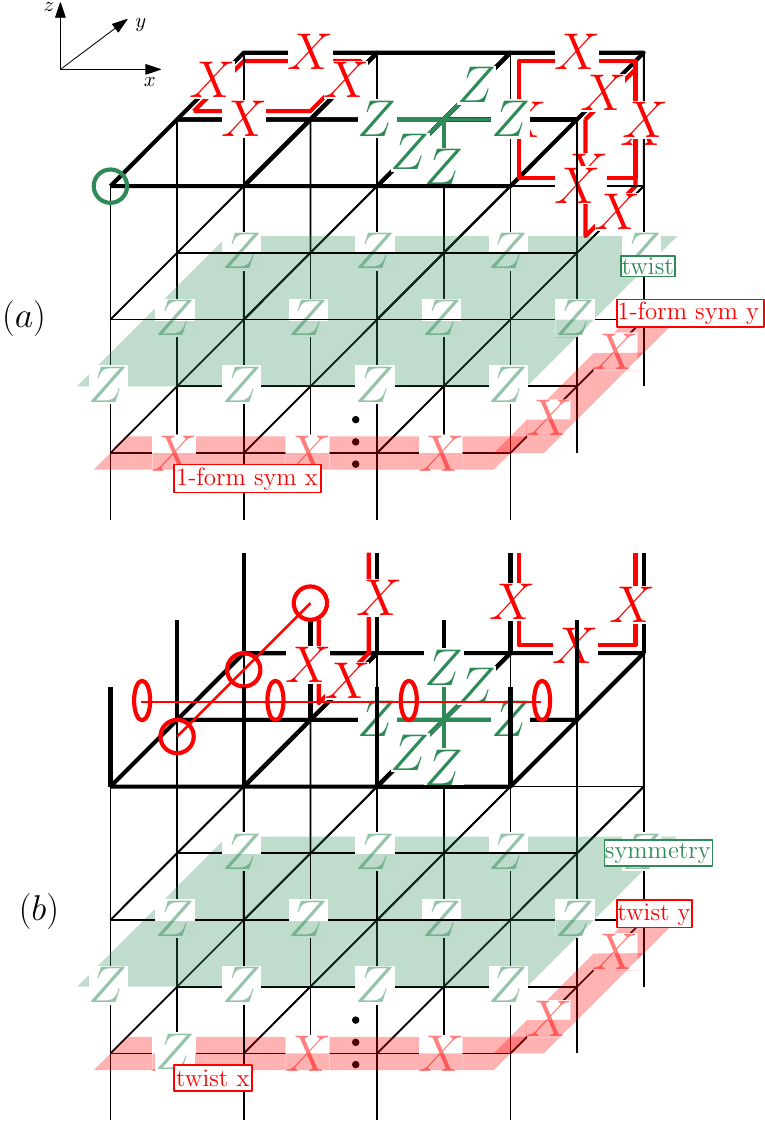}
    \caption{$3D$ toric code in the sandwich picture with smooth top boundary (a) and rough top boundary (b). Boundaries in the $x$ and $y$ direction are periodic. The unshaded terms are the boundary stabilizers (enforced) and the eigenvalues of the shaded terms label the symmetry or twist sectors depending on the top boundary. The hollow circles denote the location of omitted boundary stabilizers. For visual clarity, we have not drawn the operators $\mathcal{X}_z$, $\mathcal{Z}_{xz}$, or $\mathcal{Z}_{yx}$ which toggle between the sectors of the Hilbert space.}
    \label{3DTC}
\end{figure}

In this section, we illustrate a $2+1D$ sandwich construction with the $3+1D$ toric code in the bulk and the associated dualities implemented by two-dimensional sequential quantum circuits that map between the different gapped (topological) boundaries. We would like to mention that the generalization of the systematic construction of dualities for $1+1D$ systems in terms of module categories to $2+1D$ systems in terms of 2-categories has been suggested in Ref.~\onlinecite{Lootens2024dualities}, although the complete framework is still being developed. There, the duality is implemented by the action of a Projected Entangled Pair Operator (PEPO) on the state and the topological sectors can be obtained by using the $2D$ generalization of the tube algebra. Here, we present the duality through unitary sequential quantum circuit, in line with our claim that duality corresponds to a boundary change in the TH picture achieved with sequential quantum circuits.

The discussion is presented as a natural generalization of the $2+1D$ case. Imagine a $3+1D$ toric code model with periodic boundary conditions in the $x$ and $y$ directions (torus geometry), and a gapped boundary on top (just like in the $2D$ toric code case). The bulk Hamiltonian (in the string-net condensed picture) is
\begin{align} \label{eq:3dtc_ham}
 H_{\text{3DTC}}^{\text{bulk}} = &-\sum_{v} \left[\includegraphics[valign=c,page=31,scale=0.35]{duality_figures}\right] - \sum_{p,xz} \left[\includegraphics[valign=c,page=32,scale=0.35]{duality_figures}\right] \nonumber \\
 &- \sum_{p,yz} \left[\includegraphics[valign=c,page=33,scale=0.35]{duality_figures}\right] - \sum_{p,zy} \left[\includegraphics[valign=c,page=34,scale=0.35]{duality_figures}\right],
\end{align}
with the plaquettes living in the three planes according to Fig.~\ref{3DTC}. The bulk topological phase hosts two types of excitations: point-like charge ($\bm{e}$) excitations that live at the endpoints of $X$-strings and string-like flux-loop excitations ($\bm{m}$) that live at the boundary of membranes formed by $Z$-operators on the dual lattice. There are an infinite number of gapped boundaries, since one can create a new gapped boundary by stacking the original one with a suitable $2D$ topological phase. Our discussion is centered around three possible boundaries usually called `rough', `smooth' (in analogy with the $2D$ case), and `twisted smooth' \cite{Zhao2023,Ji2023}. These three boundaries are particularly simple since they do not contain fractional excitations confined only to the boundary. On a rough boundary, the point-like ($\bm{e}$) excitations condense and the membrane operators create string-like boundary excitations. On a smooth boundary, the flux-loops condense and the $X$-strings create point-like boundary excitations ($\bm{e}$) when ended on the boundary.  
On a `twisted smooth' boundary, the excitations that condense are the same as on the smooth boundary, but now the 
endpoints of flux loops that terminate on the boundary are attached to semionic point excitations. In other words, the two smooth boundaries differ by stacking a $2+1D$ double semion model \cite{levin2005string} on the boundary and condensing the composite of the $\bm{e}$ excitation and the boson of the double semion model, which in turn binds semions to the endpoints of flux loops. The smooth and rough boundary are shown in Fig.~\ref{3DTC} with their boundary Hamiltonian terms in the TH setup. In the twisted smooth boundary (not shown), the vertex terms on the boundary ($Z^{\otimes 5}$) get dressed by appropriate phases to recover the double semion Hamiltonian. 

We describe the various operators in the TH and their roles, as illustrated in Fig.~\ref{3DTC}. We again have local bulk (see Eq.~\ref{eq:3dtc_ham}) and boundary stabilizers whose $+1$ eigenspace defines the low-energy subspace in which the $2+1D$ theory lives. 
As in the $2D$ toric code case, we omit some boundary stabilizers to allow for twist sectors, as indicated in Fig.~\ref{3DTC}. On the rough boundary, we omit one truncated vertex term. On the smooth boundary, we omit one column and one row of truncated plaquette terms. Note that all of the omitted plaquettes in a given direction are equivalent to each other up to the enforced plaquette terms, so each direction allows for one $\pm 1$ sector label. In the bulk, there are nonlocal terms that form the logical operators of the various sectors, of which there are eight in total. The sectors are labelled by the $\pm 1$ eigenvalues of a $xy$-oriented dual plane of $Z$'s that we call $\mathcal{Z}_{xy}$ and a pair of lines of $X$'s along noncontractible loops in the $x$ and $y$ directions that we call $\mathcal{X}_x$ and $\mathcal{X}_y$. The operators that change the sectors are $z$-oriented lines of $X$'s ($\mathcal{X}_z$) and $yz$- and $xz$-oriented dual planes of $Z$'s ($\mathcal{Z}_{yz}$, $\mathcal{Z}_{xz}$), respectively. Although we do not show them explicitly, there are also local terms at the bottom boundary that commute with all the above terms and form the algebra of local symmetric operators.

With a rough top boundary, $\mathcal{X}_z$ can end anywhere without creating excitations, while $\mathcal{Z}_{xz}$ and $\mathcal{Z}_{yz}$ must end along the row or column containing the omitted stabilizers. Therefore, we identify $\mathcal{Z}_{xy}$, whose eigenvalue is toggled by $\mathcal{X}_z$, with a 0-form $\mathbb{Z}_2$ symmetry of the sandwich. The four twist sectors are then labelled by the eigenvalues of $\mathcal{X}_x$ and $\mathcal{X}_y$. Conversely, with a smooth top boundary, $\mathcal{X}_x$ and $\mathcal{X}_y$ become 1-form symmetries of the sandwich (since they are deformable in the bulk) that contribute four symmetry sectors, while the eigenvalue of $\mathcal{Z}_{xy}$ labels the two twist sectors.
Here we see an important difference with the $2D$ toric code case, since the kind of symmetry (0-form or 1-form) and the structure of the sectors changes with the choice of top boundary. Therefore, the dualities that come from changing the top boundary are dualities between theories with different types of symmetries.

We show in Table~\ref{tab:3dtc_phases} the various $2+1D$ phases that emerge in the TH picture for different choices of the top and bottom boundary of the sandwich. Dualities then correspond to interchanging the columns of this table. In particular, changing from a rough to a smooth boundary corresponds to a duality between systems with 0-form and 1-form symmetry. This is the well-known duality of gauging the $0$-form $\mathbb{Z}_2$ symmetry in $2+1D$ Ising system as discussed in Refs.~\onlinecite{kogut1979introduction, fisher2004duality, savit1980duality, wegner1971duality}. When one of the boundaries is twisted-smooth, the corresponding $2+1D$ phases can be `twisted', resulting in an SPT phase or a double semion phase, which can be regarded as a twisted toric code. The asymmetry between the twisted-smooth top boundary and rough bottom boundary versus the reversed case comes from the fact that, while both cases generate the same state of the sandwich (up to a spatial inversion), they are regarded as having different symmetries. With a rough top boundary, there is a 0-form symmetry that protects the SPT phase. With twisted-smooth top boundary, there is a 1-form symmetry which is unable to protect the SPT phase, leading to a trivial phase. We note that an explicit lattice Hamiltonian for the $2+1D$ SPT phase with a 0-form $\mathbb{Z}_2$ symmetry which also possesses a 1-form symmetry was given in~\onlinecite{Webster2022}.

\begin{table}[]
    \centering
    \begin{tabular}{|c||ccc|}
    \hline
          & \begin{tabular}{@{}c@{}} Rough \\ (0-form) \end{tabular} & \begin{tabular}{@{}c@{}} Smooth \\ (1-form) \end{tabular} & \begin{tabular}{@{}c@{}} Tw. Smooth \\ (1-form) \end{tabular} \\\hline\hline
        Rough & SSB & Trivial & Trivial \\
        Smooth & Trivial & T.C. & D.S. \\
        Tw. Smooth & SPT & D.S.& T.C. \\
        \hline
    \end{tabular}
    \caption{All $2+1D$ phases corresponding to different choices of the top boundary (columns) or bottom boundary (rows) of the $3+1D$ toric code in the TH picture. Dualities correspond to permuting columns of the table. The acronyms T.C and D.S. describe the toric code and double semion phases. The choice of top boundary also affects whether the system is regarded as having a 0-form or 1-form $\mathbb{Z}_2$ symmetry, as indicated. }
    \label{tab:3dtc_phases}
\end{table}

\subsection{Rough to smooth duality}

 \begin{figure}[b]
 \includegraphics[width=\linewidth,scale=0.1,page=57]{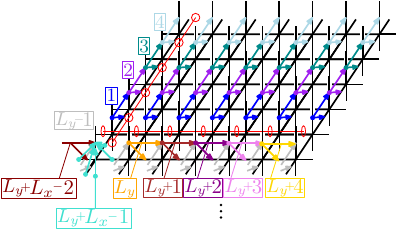}
\caption{The sequential quantum circuit that maps the rough to smooth boundary. The arrows are CNOT gates that point from control to target qubit. Gates that are applied in parallel have the same color. Hollow circles denote the omitted plaquette stabilizers of the rough houndary, and the omitted vertex stabilizer of the resulting smooth boundary is on the bottom left vertex.}
\label{3DTCseqCirc}
\end{figure}

In the same vein as the $1+1D$ case, we present a unitary sequential quantum circuit in Fig.~\ref{3DTCseqCirc} that maps the boundary from rough to smooth. As shown in the figure, in the first $L_y-1$ steps of the circuit one row of 2 CNOT gates per leg is applied in parallel (indicated by the same color). This step is linear depth in the $y$-direction. Next, a series of gates is applied sequentially on the first row in the $x$-direction. Finally, we act with more CNOTs around the bottom left vertex. The total circuit is therefore linear depth ($O(L_x+L_y-2)$). 

For every rough edge (except the bottom left one), there is a truncated plaquette term that is mapped to a single $X$ as in Eq.~\ref{rough-smooth}, which stabilizes that edge into the $|+\rangle$ state. For the rough leg in the bottom left corner, we have
\begin{align} \label{lowerVertex}
\includegraphics[valign=c,page=37,scale=0.5]{duality_figures},
\end{align}
which stabilizes that edge into the state $|0\rangle$. Finally, all other boundary vertex terms control only one CNOT gate that targets the bulk, which modifies them into the truncated vertex terms of a smooth boundary,
\begin{align} \label{3Drough-smooth}
\includegraphics[valign=c,page=35,scale=0.5]{duality_figures}.
\end{align}
Therefore, the boundary after all gates is a smooth boundary with the truncated vertex term in the bottom left corner omitted. The map from smooth to rough can be accomplished by adding ancilla qubits in the correct state ($|+\rangle$ or $|0\rangle$) and running the same circuit in reverse.

From Table \ref{tab:3dtc_phases}, we see that the rough to smooth duality corresponds to the usual ``gauging'' duality that maps a trivial (SPT) phase to a toric code (double semion) phase and maps a symmetry-breaking phase to a ``Higgsed'' trivial phase. When the boundary changes, the roles of the bulk string and plane operators as labelling symmetry or twist sectors are exchanged. The label for the $x$ and $y$ twist sectors for the rough boundary become the $x$ and $y$ 1-form symmetry sectors, and the symmetry sector becomes the twist sector. Just like in the $2D$ toric code case, there is a one-to-one correspondence between bulk and boundary sectors and therefore there is no requirement for ancilla qubits in the circuit in Fig. \ref{3DTCseqCirc}.

As with the Kramers-Wannier duality (Sec.~\ref{sec:KW}), because the duality is implemented by a unitary operator, we should be able to explicitly match the low-energy spectra on each side of this $2+1D$ gauging duality. We consider fixed-point models on a square lattice with $N$ sites and periodic boundary conditions. The duality then maps between a system with qubits on the sites of the square lattice with the 0-form symmetry $\prod_i X_i$ to a model with qubits on the edges of the square lattice with the 1-form symmetry $\prod_{e\in\gamma} Z_e$ where $\gamma$ is any closed path of edges. 

Let us first consider the trivial symmetry and twist sectors. In the trivial phase with 0-form $\mathbb{Z}_2$ symmetry, the ground state is a unique paramagnetic state $|+\rangle^{\otimes N}$ and the first excitations are created by flipping $|+\rangle \rightarrow |-\rangle$ on any two sites. In the dual toric code phase, the ground state is also unique because the 1-form symmetry selects a particular toric code ground state on any spatial manifold (the $+1$ eigenstate of non-contractible $m$-loops in both directions). The first excitations are $e$-anyons, which live on sites and necessarily come in pairs (the $m$-anyons do not appear as excitations due to the enforced 1-form symmetry), such that they have the same energy and multiplicity as the excitations in the trivial phase. A similar discussion holds for the duality between the SPT and the double-semion phases. For the duality between the symmetry-breaking and the ``Higgsed'' trivial phases, the symmetry-breaking state has a unique symmetric ground state (the superposition of the two symmetry-breaking states), as does the Higgsed phase (the 1-form symmetric product state $|0\rangle^{\otimes N}$). In the symmetry-breaking phase, excitations are domain walls which form closed-loops on the dual-lattice. In the Higgsed phase, excitations are spin flips which must also form closed loops on the dual lattice due to the 1-form symmetry. 

We can similarly match the spectra between non-trivial symmetry or twist sectors on either side of the duality. For instance, the $-1$ symmetry sector of the trivial phase with 0-form symmetry maps to the $-1$ twist sector of the toric code. The former is spanned by product states with an odd number of sites in the $|-\rangle$ state, while the latter are eigenstates of a Toric Code Hamiltonian with the sign of one plaquette term flipped. In each case, there are $N$ ground states and $\binom{N}{3}$ first excited states. In the symmetry-breaking phase with 0-form symmetry, a twist in the $x$ ($y$) direction is obtained by flipping the sign of Ising terms along a non-contractible path of edges wrapping in the $y$ ($x$) direction. This maps to states in the Higgsed phase that have an odd number of flipped spins along every non-contractible path on the dual lattice in the $x$ ($y$) direction. On an $L_x\times L_y$ torus, there are $L_y$ ($L_x$) ground states and $L_xL_y$ first excited states in both phases.

\subsection{Smooth to twisted-smooth duality}

Now we discuss the duality that is induced by changing from a smooth boundary to a twisted-smooth boundary. First, we describe a simple way to construct the twisted smooth boundary. The $3+1D$ toric code can be obtained via a gauging procedure which maps short-range entangled states with a 0-form $\mathbb{Z}_2$ symmetry to long-range entangled states with a 1-form $\mathbb{Z}_2$ symmetry \cite{Ji2023}. In particular, if we begin with a trivial $3+1D$ $\mathbb{Z}_2$-symmetric state, then the gauging procedure maps this to the $3+1D$ toric code. To construct the toric code with boundary, we simply truncate the gauging procedure to a finite volume of space. At the boundary of this volume, we have the freedom to put any $2+1D$ $\mathbb{Z}_2$ symmetric state before gauging. The different choices of this boundary state lead to the different gapped boundaries after gauging. Specifically, starting with a $2+1D$ boundary state belonging to the trivial, symmetry-breaking, or SPT phase with $\mathbb{Z}_2$ symmetry leads to the smooth, rough, and twisted-smooth boundaries after gauging, respectively \cite{Ji2023}. 

The perspective on gapped boundaries in terms of gauging also allows us to construct SQCs that change boundary type. For example, to construct an SQC that changes from the smooth to the twisted-smooth boundary, we can start with an SQC that maps from the trivial $\mathbb{Z}_2$-symmetric $2+1D$ state to an SPT-ordered state such as the Levin-Gu state \cite{Levin2012}. Such an SQC was constructed in Ref.~\onlinecite{Chen2024}. Then, we can pull this SQC through the gauging map discussed above to obtain a new SQC that maps from the smooth to the twisted-smooth boundary. More specifically, assuming that the $\mathbb{Z}_2$ symmetry corresponds to applying $X$ to all lattice sites, this `gauged' SQC acts on a Hilbert space consisting of qubits on the edges of the lattice, and is obtained from the original SQC via a transformation of the Pauli operators where $X$ on a site is mapped to the product of $X$'s on all incident edges, and a pair of $Z$'s on two sites are mapped to a string of $Z$'s on edges connecting the two sites, see Refs.~\onlinecite{Williamson2016,Shirley2019} for details. 

Crucially, the gauged SQC still consists of local unitary gates since the gauging map preserves unitarity and the original SQC consists of $\mathbb{Z}_2$ symmetric gates. The latter fact is important because, while the gauging map is not locality preserving in general, it does map $\mathbb{Z}_2$-symmetric local operators to local operators. Therefore, gauging takes each local unitary gate in the SQC to another local unitary gate, resulting in a gauged SQC with the same structure as the original one that maps between the smooth and twisted-smooth boundaries. We remark that a similar method could have been used to cosntruct the SQC mapping from rough to smooth boundary. Indeed, the circuit in Fig.~\ref{3DTCseqCirc} strongly resembles the SQC used to map between the $2+1D$ symmetric and symmetry-breaking phases constructed in Ref.~\cite{Chen2024}.

We now describe the $2+1D$ duality implemented by changing the boundary from smooth to twisted-smooth in the TH picture. An important distinction from the case of the rough-to-smooth duality in $3+1D$ toric codes is that the symmetry is the same on both sides of the duality: a $\mathbb{Z}_2$ 1-form symmetry. In terms of the permutation of phase diagrams, Table \ref{tab:3dtc_phases} shows that the duality corresponds to exchanging the toric code and double semion phases in $2+1D$ while preserving the trivial phase. As required by spectrum matching, the ground state degeneracy and number of distinct anyons in the toric code and double semion models are the same.

\section{summary}
\label{sec:summary}

In this paper, we discuss duality in the Topological Holography / Symmetry TFT / Symmetry TO framework where duality is induced by a change in the top boundary condition of a sandwich structure. We focus on 1. the $1+1D$ Kramers-Wannier duality realized in a sandwich structure with the $2+1D$ Toric Code in the bulk; 2. general $1+1D$ dualities with $2+1D$ string-nets in the bulk and in particular the $S_3$ quantum double giving rise to a non-invertible symmetry in the sandwich; 3. the $2+1D$ gauging duality realized with $3+1D$ Toric Code in the bulk where such that the symmetry of the sandwich transform between a $0$-form symmetry and a $1$-form symmetry. We demonstrate, in particular, that duality can be realized with a Sequential Quantum Circuit applied to the top boundary only without affecting the bulk or the bottom boundary. As a consequence, the Hamiltonians before and after the duality mapping have
the same spectrum in the corresponding symmetry sectors (up to exact degeneracy in some cases), and the entanglement in the
corresponding low-energy eigenstates differ by at most an area law term. 

The Topological Holography framework provides a very concrete set up to discuss duality, so we can discuss explicitly what is kept invariant and what can change under the duality transformation. Let's distinguish between a few concepts: symmetry (non-generalized), duality symmetry and (general) duality.
\begin{enumerate}
\item When we say a system has a symmetry (in the conventional sense), we mean the Hamiltonian remains invariant under the symmetry transformation. So the quantum system before and after the transformation is the same (with the same DOF, dimension, and symmetry), the phases are the same, the spectrum is the same, and the eigenstates form representations of the symmetry.
\item When a system has a duality symmetry, under the duality transformation, the system is still the same system (with the same DOF, dimension, and symmetry), but the phases are permuted. More importantly, the symmetry sectors are permuted, so the charge sector can map to the flux sector for example. This is achieved in the sandwich picture by changing the top boundary such that the symmetry of the sandwich remains the same but realized differently. The duality symmetry induces a locality preserving operator mapping (Quantum Cellular Automaton) on the symmetric operator as they live at the bottom boundary and are not affected by the duality transformation (although their interpretation as symmetric Hamiltonian terms can change).
\item The more general version of duality discussed in our paper can map one quantum system to another system (bosonic to fermionic, $S_3$ symmetry to Rep($S_3$) symmetry, $0$-form symmetry to $1$-form symmetry, etc). This is realized in the sandwich picture with a generic change of top boundary. It induces a map from the phases and phase transitions of system 1 to that of system 2. The symmetry and twist sectors of the two systems have a fixed correspondence, such that before and after the duality mapping, the spectrum in corresponding sectors match with each other (up to exact degeneracies). 
\item There is an interesting case in between 1 and 2. It is realized in the sandwich picture by applying a bulk symmetry to the whole sandwich such that the top boundary remains invariant. Under this mapping, the sandwich still represents the same quantum system with the same DOF, dimension, and symmetry while the phase can be permuted. The difference from 2 is that, because the top boundary remains invariant, the symmetry sectors also remain invariant. Therefore, for the phases before and after such a duality mapping, the spectrum in each sector remains invariant. In this case, the mapping induces a Quantum Cellular Automaton for the full operator algebra, not just for the symmetric operators as in case 2. An example of this type exists in the $2+1D$ sandwich with the $3+1D$ Toric Code bulk. The bulk has a $\mathbb{Z}_2$ symmetry that changes the smooth boundary to the twisted smooth boundary while keeping the rough boundary invariant. Therefore, when the top boundary is set to be rough, the bulk symmetry exchanges the trivial and nontrivial SPT phases with $0$-form $\mathbb{Z}_2$ symmetry. 
\end{enumerate}

An interesting open question is how much the formalism developed in this paper applies to the particle-vortex duality (or the related ones in the duality web). The basic version of TH works for finite (invertible or noninvertible) symmetries, but it has recently been generalized  to continuous symmetries as well \cite{brennan2024symtft, antinucci2024arxiv}. In Ref.~\onlinecite{Seiberg2016}, it was shown how the $2+1D$ particle-vortex duality is related to the electric-magnetic duality in the bulk $3+1D$ $U(1)$ gauge theory. We will leave the discussion of particle-vortex duality and other dualities in systems with continuous symmetries to the future.


\begin{acknowledgments}
We are indebted to inspiring discussions with Michael Levin, Liang Kong, Frank Verstraete, Laurens Lootens, Dominic Williamson and Wenjie Ji. The authors acknowledge support from the Simons collaboration on ``Ultra-Quantum Matter'' [Grant No. 651438 (X.C.) and 651440 (D.T.S.)], the Simons Investigator Award (X.C. and R.V. award ID 828078). X.C. is supported by the Walter Burke Institute for Theoretical Physics at Caltech and the Institute for Quantum Information and Matter at Caltech. R.V. is supported by the Research Foundation Flanders (FWO).
X.G.W is partially supported by NSF DMR-2022428 and by the Simons 
Collaboration on Ultra-Quantum Matter [Grant No. 651446, X.G.W]
\end{acknowledgments}
\newpage

\bibliography{references}

\appendix

\section{$S_3$ quantum double sector identification}
\label{S3data}

In this appendix, we show how the eight bulk sectors for the $S_3$ quantum double decompose into boundary sectors in different ways for the gapped boundaries $\mathcal{A}_1$ and $\mathcal{A}_4$ (table \ref{anyonsS3}).
 
We consider the eight topological sectors labeled by $\bm{A},\bm{B},\bm{C},\dots,\bm{H}$. A sector $\beta$ is fixed by applying the corresponding projector in the horizontal direction of the sandwich (Eq.~\ref{HilbertSpaceBeta}). Using the $S$-matrix for the $S_3$ quantum double (see for example Ref.~\cite{beigi2011quantum}), we can write:
\begin{align}
\label{S3sectors}
\begin{split}
P_{\bm{A}} &=  \frac{1}{36}(\mathcal{L}_{\bm{A}}+\mathcal{L}_{\bm{B}}+2\mathcal{L}_{\bm{C}}+2\mathcal{L}_{\bm{F}}+2\mathcal{L}_{\bm{G}}+2\mathcal{L}_{\bm{H}}+\\
&\quad\quad\quad 3\mathcal{L}_{\bm{D}}+3\mathcal{L}_{\bm{E}})\\
P_{\bm{B}} &=  \frac{1}{36}(\mathcal{L}_{\bm{A}}+\mathcal{L}_{\bm{B}}+2\mathcal{L}_{\bm{C}}+2\mathcal{L}_{\bm{F}}+2\mathcal{L}_{\bm{G}}+2\mathcal{L}_{\bm{H}}-\\
&\quad\quad\quad 3\mathcal{L}_{\bm{D}}-3\mathcal{L}_{\bm{E}})\\
P_{\bm{C}} &= \frac{1}{9}(\mathcal{L}_{\bm{A}}+\mathcal{L}_{\bm{B}}+2\mathcal{L}_{\bm{C}}-\mathcal{L}_{\bm{F}}-\mathcal{L}_{\bm{G}}-\mathcal{L}_{\bm{H}})\\
P_{\bm{D}} &= \frac{1}{4}(\mathcal{L}_{\bm{A}}-\mathcal{L}_{\bm{B}}+\mathcal{L}_{\bm{D}}-\mathcal{L}_{\bm{E}})\\
P_{\bm{E}} &= \frac{1}{4}(\mathcal{L}_{\bm{A}}-\mathcal{L}_{\bm{B}}-\mathcal{L}_{\bm{D}}+\mathcal{L}_{\bm{E}})\\
P_{\bm{F}} &=  \frac{1}{9}(\mathcal{L}_{\bm{A}}+\mathcal{L}_{\bm{B}}-\mathcal{L}_{\bm{C}}+2\mathcal{L}_{\bm{F}}-\mathcal{L}_{\bm{G}}-\mathcal{L}_{\bm{H}})\\
P_{\bm{G}} &=  \frac{1}{9}(\mathcal{L}_{\bm{A}}+\mathcal{L}_{\bm{B}}-\mathcal{L}_{\bm{C}}-\mathcal{L}_{\bm{F}}+2\mathcal{L}_{\bm{G}}-\mathcal{L}_{\bm{H}})\\
P_{\bm{H}} &=  \frac{1}{9}(\mathcal{L}_{\bm{A}}+\mathcal{L}_{\bm{B}}-\mathcal{L}_{\bm{C}}-\mathcal{L}_{\bm{F}}-\mathcal{L}_{\bm{G}}+2\mathcal{L}_{\bm{H}}).
\end{split}
\end{align}

The bulk Wilson lines $\mathcal{L}$ interact non-trivially with the boundary when they are brought close to it. They decompose into `boundary Wilson lines' in the following way: 

\begin{align}
\label{boundarySplitting}
\begin{split}
\mathcal{L}_{\alpha} &= \sum_{a,i,j} \mathcal{L}_{\alpha}^{a,i,j}\\
\mathcal{L}_{\alpha}^{a,i,j} &= \  \includegraphics[valign=c,page=61,scale=0.6]{duality_figures}, \ a \in \mathcal{C}_{\mathcal{M}}^*, 
\end{split}
\end{align}
where we have used a different normalization compared to Ref.~\onlinecite{lin2023asymptotic}. Just like the bulk Wilson lines can be interpreted as an anyon pair creation and annihilation process, the boundary Wilson lines can be interpreted as a process where a bulk anyon $\alpha$ is created from the boundary, leaving behind an excitation $a$, after which it is wrapped around the cylinder and absorbed again on the boundary, annihilating the boundary excitation. Using suitable boundary recouplings, the boundary Wilson lines can be formulated as bulk Wilson lines, connected to the boundary by a vertical condensed Wilson line. Making this correspondence is not the aim of this discussion and we will write the boundary sectors in terms of the boundary Wilson lines in Eq. \ref{boundarySplitting}.

To decompose the bulk sectors into boundary sectors, labeled by simple boundary excitations $b$ and a possible degeneracy, we have to project them on the subspaces of the Hilbert space that correspond to simple boundary excitations. These subspaces are characterized by the eigenvalues of the condensed anyons, connected to the boundary $\bm{\mu}_{i,j}, \ \bm{\mu} \in \mathcal{A}$:
\begin{align} 
\begin{split}
\mathcal{H}^{b, \mathcal{M}(\mathcal{A})} &= P^b \ \mathcal{H}^{\mathcal{M}(\mathcal{A})} \\
P^b &=  \sum_{\mu \in \mathcal{A},i,j} \gamma_{1}^b \ \bar{\gamma}_{\mu,(i,j)}^{b} \
\mathcal{L}_{\mu}^{i,j} \ ,
\end{split}
\label{HilbertSpaceb}
\end{align}
with $\mathcal{L}_{\mu}^{i,j} = \mathcal{L}_{\mu}^{1,i,j}$ (since $\mu$ is condensed). The matrix $\gamma$ is the half-braiding matrix (called the half-linking in Ref.~\onlinecite{shen2019defect, lin2023asymptotic}) and is a subset of what is generally known as the half-braiding tensor (denoted by $\Omega$) \cite{levin2005string}, which can be (loosely) considered as the generalization of the bulk $S$ matrix on the boundary, in the sense that it performs the basis change between the boundary Wilson lines and the boundary sectors. The bulk sectors split into non-vanishing boundary sectors $P_{\beta}^{b,i}$, where i labels a possible degeneracy if $W_{\beta}^b>1$. The evaluation of the action of $P^b$ on a bulk sector can be performed by establishing a multiplication between boundary Wilson lines. This can be done by explicitly writing the action of the boundary Wilson lines on the string-net with boundary and see how they multiply. In this way, the boundary Wilson lines form an algebra, equivalent to the tube algebra (after a basis change) (see for example Ref.\cite{williamson2017symmetry}). We won't do the explicit calculation here, but will simply show the result on the decomposition of the sectors. \\

A smooth (charge condensed or canonical) boundary is characterized by the condensable Lagrangian algebra $\mathcal{A}_1 = \bm{A}+\bm{B}+2\bm{C}$ and the symmetry of the sandwich is $\text{Vec}_{S_3}$ (see table \ref{anyonsS3}). The Wilson lines $\mathcal{L}_{\bm{A}}$ and $\mathcal{L}_{\bm{B}}$ trivially connect to the boundary. The anyon $\bm{C}$ has 2 condensation channels and the corresponding Wilson line decomposes as $\mathcal{L}_{\bm{C}} = \mathcal{L}_{\bm{C}}^{1,1} + \mathcal{L}_{\bm{C}}^{1,2} + \mathcal{L}_{\bm{C}}^{2,1} + \mathcal{L}_{\bm{C}}^{2,2}$. The anyons $\bm{F}$, $\bm{G}$ and $\bm{H}$ are identified under condensation and split into two blocks on the boundary (\ref{boundarySplitting}). The Wilson lines decompose as $\mathcal{L}_{\bm{F}}=\mathcal{L}_{\bm{F}}^{r}+\mathcal{L}_{\bm{F}}^{\bar{r}}$, $\mathcal{L}_{\bm{G}}=\mathcal{L}_{\bm{G}}^{r}+\mathcal{L}_{\bm{G}}^{\bar{r}}$, $\mathcal{L}_{\bm{H}}=\mathcal{L}_{\bm{H}}^{r}+\mathcal{L}_{\bm{H}}^{\bar{r}}$. Similar for the anyons $\bm{D}$ and $\bm{E}$. The Wilson lines decompose into three blocks on the boundary: $\mathcal{L}_{\bm{D}} = \mathcal{L}_{\bm{D}}^{s} + \mathcal{L}_{\bm{D}}^{sr} + \mathcal{L}_{\bm{D}}^{s\bar{r}}$, $\mathcal{L}_{\bm{E}} = \mathcal{L}_{\bm{E}}^{s} + \mathcal{L}_{\bm{E}}^{sr} + \mathcal{L}_{\bm{E}}^{s\bar{r}}$. 

The 6 projectors on the simple boundary excitations are (Eq.~\ref{HilbertSpaceb}):
\begin{align*}
\begin{split}
P^1 &= \frac{1}{6}(\mathcal{L}_{\bm{A}}+\mathcal{L}_{\bm{B}}+2\mathcal{L}_{\bm{C}}^{1,1}+2\mathcal{L}_{\bm{C}}^{2,2})\\
P^r &= \frac{1}{6}(\mathcal{L}_{\bm{A}}+\mathcal{L}_{\bm{B}}+2\omega\mathcal{L}_{\bm{C}}^{1,1}+2\bar{\omega}\mathcal{L}_{\bm{C}}^{2,2})\\
P^{\bar{r}} &= \frac{1}{6}(\mathcal{L}_{\bm{A}}+\mathcal{L}_{\bm{B}}+2\bar{\omega}\mathcal{L}_{\bm{C}}^{1,1}+2\omega\mathcal{L}_{\bm{C}}^{2,2})\\
P^s &= \frac{1}{6}(\mathcal{L}_{\bm{A}}-\mathcal{L}_{\bm{B}}+2\mathcal{L}_{\bm{C}}^{1,2}+2\mathcal{L}_{\bm{C}}^{2,1})\\
P^{sr} &= \frac{1}{6}(\mathcal{L}_{\bm{A}}-\mathcal{L}_{\bm{B}}+2\omega\mathcal{L}_{\bm{C}}^{1,2}+2\bar{\omega}\mathcal{L}_{\bm{C}}^{2,1})\\
P^{s\bar{r}} &= \frac{1}{6}(\mathcal{L}_{\bm{A}}-\mathcal{L}_{\bm{B}}+2\bar{\omega}\mathcal{L}_{\bm{C}}^{1,2}+2\omega\mathcal{L}_{\bm{C}}^{2,1}).
\end{split}
\end{align*}
Note that we have used a different basis for the half-linking matrix compared to Ref.~\onlinecite{shen2019defect}. The eight bulk sectors split into 16 boundary sectors in the following way:
\begin{align}
\label{sectorsplittingA1}
\begin{split}
P_{\bm{A}} &= P_{\bm{A}}^1\\
P_{\bm{B}} &= P_{\bm{B}}^1\\
P_{\bm{C}} &= P_{\bm{C}}^{1,1,1} + P_{\bm{C}}^{1,1,2}\\
P_{\bm{D}} &= P_{\bm{D}}^{s} + P_{\bm{D}}^{sr} + P_{\bm{D}}^{s\bar{r}}\\
P_{\bm{E}} &= P_{\bm{E}}^{s} + P_{\bm{E}}^{sr} + P_{\bm{E}}^{s\bar{r}}\\
P_{\bm{F}} &= P_{\bm{F}}^{r} +P_{\bm{F}}^{\bar{r}} \\
P_{\bm{G}} &= P_{\bm{G}}^{r} +P_{\bm{G}}^{\bar{r}} \\
P_{\bm{H}} &= P_{\bm{H}}^{r} +P_{\bm{H}}^{\bar{r}} \\
\end{split}
\end{align}

with the individual boundary sectors expressed in terms of the boundary Wilson lines as:
\begin{widetext}
\begin{align*}
\begin{split}
P_{\bm{C}}^{1,1,1} &= \frac{1}{18}(\mathcal{L}_{\bm{A}}+\mathcal{L}_{\bm{B}}+2\mathcal{L}_{\bm{C}}+2\omega(\mathcal{L}_{\bm{F}}^r+\mathcal{L}_{\bm{G}}^r+\mathcal{L}_{\bm{H}}^r) +2\bar{\omega}(\mathcal{L}_{\bm{F}}^{\bar{r}}+\mathcal{L}_{\bm{G}}^{\bar{r}}+\mathcal{L}_{\bm{H}}^{\bar{r}})) \\
P_{\bm{C}}^{1,1,2} &= \frac{1}{18}(\mathcal{L}_{\bm{A}}+\mathcal{L}_{\bm{B}}+2\mathcal{L}_{\bm{C}}+2\bar{\omega}(\mathcal{L}_{\bm{F}}^r+\mathcal{L}_{\bm{G}}^r+\mathcal{L}_{\bm{H}}^r) +2\omega(\mathcal{L}_{\bm{F}}^{\bar{r}}+\mathcal{L}_{\bm{G}}^{\bar{r}}+\mathcal{L}_{\bm{H}}^{\bar{r}})) \\
P_{\bm{D}}^{s} &= \frac{1}{12}(\mathcal{L}_{\bm{A}}-\mathcal{L}_{\bm{B}}+6\mathcal{L}_{\bm{C}}^{1,2}+6\mathcal{L}_{\bm{C}}^{2,1} + 3\mathcal{L}_{\bm{D}}^{s}-3\mathcal{L}_{\bm{E}}^{s}) \\
P_{\bm{D}}^{sr} &= \frac{1}{12}(\mathcal{L}_{\bm{A}}-\mathcal{L}_{\bm{B}}+6\omega\mathcal{L}_{\bm{C}}^{1,2}+6\bar{\omega}\mathcal{L}_{\bm{C}}^{2,1} + 3\mathcal{L}_{\bm{D}}^{s}-3\mathcal{L}_{\bm{E}}^{s}) \\
P_{\bm{D}}^{s\bar{r}} &= \frac{1}{12}(\mathcal{L}_{\bm{A}}-\mathcal{L}_{\bm{B}}+6\bar{\omega}\mathcal{L}_{\bm{C}}^{1,2}+6\omega\mathcal{L}_{\bm{C}}^{2,1} + 3\mathcal{L}_{\bm{D}}^{s}-3\mathcal{L}_{\bm{E}}^{s}) \\
P_{\bm{F}}^{r} &= \frac{1}{18}(\mathcal{L}_{\bm{A}}+\mathcal{L}_{\bm{B}}+2\omega\mathcal{L}_{\bm{C}}^{1,1}+2\bar{\omega}\mathcal{L}_{\bm{C}}^{2,2} + 2\mathcal{L}_{\bm{F}}+2\omega\mathcal{L}_{\bm{G}}+2\bar{\omega}\mathcal{L}_{\bm{G}})\\
P_{\bm{F}}^{\bar{r}} &= \frac{1}{18}(\mathcal{L}_{\bm{A}}+\mathcal{L}_{\bm{B}}+2\bar{\omega}\mathcal{L}_{\bm{C}}^{1,1}+2\omega\mathcal{L}_{\bm{C}}^{2,2} + 2\mathcal{L}_{\bm{F}}+2\bar{\omega}\mathcal{L}_{\bm{G}}+2\omega\mathcal{L}_{\bm{G}})\\
\end{split}
\end{align*}
\end{widetext}
and with $P_{\bm{E}}^b = P_{\bm{D}}^b(\bm{E} \leftrightarrow \bm{D})$, $P_{\bm{G}}^b = P_{\bm{F}}^b(\bm{F} \leftrightarrow \bm{G})$ and $P_{\bm{H}}^b = P_{\bm{F}}^b(\bm{F} \leftrightarrow \bm{H})$. \\

For a flux-condensed boundary ($\mathcal{A}_4$ in table \ref{S3data}), $\bm{F}$ and $\bm{D}$ are partially condensed. The bulk anyons that split on the boundary split into elements of $\text{Rep}(S_3)$ (Eq.~\ref{boundarySplitting}). The bulk Wilson lines decompose as:
\begin{align*}
\mathcal{L}_{\bm{B}} &= \mathcal{L}_{\bm{B}}^{\rho_-} \\
\mathcal{L}_{\bm{C}} &= \mathcal{L}_{\bm{C}}^{\rho_2} \\
\mathcal{L}_{\bm{G}} &= \mathcal{L}_{\bm{G}}^{\rho_2} \\
\mathcal{L}_{\bm{H}} &= \mathcal{L}_{\bm{H}}^{\rho_2} \\
\mathcal{L}_{\bm{F}} &= \mathcal{L}_{\bm{F}}^{1} + \mathcal{L}_{\bm{F}}^{\rho_-}\\
\mathcal{L}_{\bm{D}} &= \mathcal{L}_{\bm{F}}^{1} + \mathcal{L}_{\bm{F}}^{\rho_2}\\
\mathcal{L}_{\bm{E}} &= \mathcal{L}_{\bm{E}}^{\rho_-} + \mathcal{L}_{\bm{E}}^{\rho_2}.
\end{align*}
Using the half-linking matrix for this boundary from Ref.~\onlinecite{shen2019defect} with a slightly different normalization, we can write down the three projectors on the subspaces corresponding to simple boundary excitations:
\begin{align*}
P^{1} &=  \frac{1}{6}(\mathcal{L}_{\bm{A}}+2 \mathcal{L}_{\bm{F}}^1+3\mathcal{L}_{\bm{D}}^1)\\
P^{\rho_-} &=  \frac{1}{6}(\mathcal{L}_{\bm{A}}+2 \mathcal{L}_{\bm{F}}^1-3\mathcal{L}_{\bm{D}}^1)\\
P^{\rho_2} &=  \frac{2}{3}(\mathcal{L}_{\bm{A}}-\mathcal{L}_{\bm{F}}^1).
\end{align*}
The eight bulk sectors split into 11 boundary sectors as (there are no degeneracies within each sector labeled by a boundary excitation $b$):
\begin{align}
\label{boundarySectorsA4}
\begin{split}
P_{\bm{A}} &= P_{\bm{A}}^1\\
P_{\bm{B}} &= P_{\bm{B}}^{\rho_-}\\
P_{\bm{C}} &= P_{\bm{C}}^{\rho_2}\\
P_{\bm{D}} &= P_{\bm{D}}^{1} + P_{\bm{D}}^{\rho_2}\\
P_{\bm{E}} &= P_{\bm{E}}^{\rho_-} + P_{\bm{E}}^{\rho_2}\\
P_{\bm{F}} &= P_{\bm{F}}^{1} + P_{\bm{F}}^{\rho_-}\\
P_{\bm{G}} &= P_{\bm{G}}^{\rho_2} \\
P_{\bm{H}} &= P_{\bm{H}}^{\rho_2}
\end{split}
\end{align}
The Hilbert space that corresponds to a non-trivial twist $\rho_2$ splits into five charge sectors in $\bm{D}, \bm{E}, \bm{C}, \bm{G}$ and $\bm{H}$, which correspond to the idempotents of the 5 elements in the tube algebra with a $\rho_2$-twist in Ref.~\onlinecite{Lootens2024dualities}. The individual boundary sectors expressed in terms of the boundary Wilson lines are:
\begin{widetext}
\begin{align}
\begin{split}
P_{\bm{D}}^{1} &= \frac{1}{12}(\mathcal{L}_{\bm{A}}-\mathcal{L}_{\bm{B}}^{\rho_-}+2\mathcal{L}_{\bm{F}}^1-2\mathcal{L}_{\bm{F}}^{\rho_-} +3\mathcal{L}_{\bm{D}}^1-3\mathcal{L}_{\bm{E}}^{\rho_-})\\
P_{\bm{D}}^{\rho_2} &= \frac{1}{12}(2\mathcal{L}_{\bm{A}}-2\mathcal{L}_{\bm{B}}^{\rho_-}-2\mathcal{L}_{\bm{F}}^1+2\mathcal{L}_{\bm{F}}^{\rho_-} +3\mathcal{L}_{\bm{D}}^{\rho_2}-3\mathcal{L}_{\bm{E}}^{\rho_2})\\
P_{\bm{E}}^{\rho_-} &=\frac{1}{12}(\mathcal{L}_{\bm{A}}-\mathcal{L}_{\bm{B}}^{\rho_-}+2\mathcal{L}_{\bm{F}}^1-2\mathcal{L}_{\bm{F}}^{\rho_-} -3\mathcal{L}_{\bm{D}}^1+3\mathcal{L}_{\bm{E}}^{\rho_-})\\
P_{\bm{E}}^{\rho_2} &= \frac{1}{12}(2\mathcal{L}_{\bm{A}}-2\mathcal{L}_{\bm{B}}^{\rho_-}-2\mathcal{L}_{\bm{F}}^1+2\mathcal{L}_{\bm{F}}^{\rho_-} -3\mathcal{L}_{\bm{D}}^{\rho_2}+3\mathcal{L}_{\bm{E}}^{\rho_2})\\
P_{\bm{F}}^{1} &= \frac{1}{18}(\mathcal{L}_{\bm{A}}+\mathcal{L}_{\bm{B}}^{\rho_-}+2\mathcal{L}_{\bm{F}}-\mathcal{L}_{\bm{C}}^{\rho_2}-\mathcal{L}_{\bm{G}}^{\rho_2}-\mathcal{L}_{\bm{H}}^{\rho_2}+3\mathcal{L}_{\bm{D}}^1-\frac{3}{2}\mathcal{L}_{\bm{D}}^{\rho_2}+3\mathcal{L}_{\bm{E}}^{\rho_-}-\frac{3}{2}\mathcal{L}_{\bm{E}}^{\rho_2})\\
P_{\bm{F}}^{\rho_-} &= \frac{1}{18}(\mathcal{L}_{\bm{A}}+\mathcal{L}_{\bm{B}}^{\rho_-}+2\mathcal{L}_{\bm{F}}-\mathcal{L}_{\bm{C}}^{\rho_2}-\mathcal{L}_{\bm{G}}^{\rho_2}-\mathcal{L}_{\bm{H}}^{\rho_2}-3\mathcal{L}_{\bm{D}}^1+\frac{3}{2}\mathcal{L}_{\bm{D}}^{\rho_2}-3\mathcal{L}_{\bm{E}}^{\rho_-}+\frac{3}{2}\mathcal{L}_{\bm{E}}^{\rho_2}).
\end{split}
\end{align}
\end{widetext}

\end{document}